\documentstyle[12pt]{article}
\newcommand{\newsection}{    % Numeration of eqs. is automatic
\setcounter{equation}{0}
\section}
\def\appendix#1{
\addtocounter{section}{1}
\setcounter{equation}{0}
\renewcommand{\thesection}{\Alph{section}}
\section*{Appendix \thesection\protect\indent #1}
\addcontentsline{toc}{section}{Appendix \thesection\ \ \ #1}
}
\newcommand{\rf}[1]{(\ref{#1})}

\def\be{\begin{equation}}
\def\ee{\end{equation}}
\newcommand{\beq}{\begin{equation}}
\newcommand{\eeq}{\end{equation}}
\newcommand{\bea}{\begin{eqnarray}}
\newcommand{\eea}{\end{eqnarray}}

\renewcommand{\l}{\lambda}
\renewcommand{\b}{\beta}
\renewcommand{\a}{\alpha}

\newcommand{\om}{\omega}

\newcommand{\non}{\nonumber}
\newcommand{\ltr}{{\,\rm Tr}\:}

\newcommand{\La}{\Lambda}

\newcommand{\eps}{\varepsilon}

\newcommand{\D}{{{\hbox{d}}}}

\newcommand{\hs}{\hspace{0.7cm}}
\begin{document}
\topmargin 0pt
\oddsidemargin 5mm
\headheight 0pt
\headsep 0pt
\topskip 9mm

\hfill SPhT/95-068
\addtolength{\baselineskip}{0.20\baselineskip}
\begin{center}
\vspace{26pt}
{\large \bf Exact solution of the O(n) model on a random lattice}
\vspace{26pt}
\newline
B.\ Eynard and C.\ Kristjansen
\footnote{On leave of absence from NORDITA, Blegdamsvej 17,
DK-2100 Copenhagen \O, Denmark} \\
\vspace{6pt}
Service de Physique Th\'{e}orique de Saclay \\
F-91191 Gif-sur-Yvette Cedex, France \\
\end{center}
\vspace{20pt}
\begin{center}
{\bf Abstract}
\end{center}
We present an exact solution of the $O(n)$ model on a random lattice.
The coupling constant space of our model is parametrized in terms of a set of
moment variables and the same type of universality with respect to the
potential as observed for the one-matrix model is found.
In addition we find a large degree of universality with respect to $n$;
namely for $n\in ]-2,2[$ the solution can be presented in a form which is valid
not only for any potential, but also for any $n$ (not necessarily rational).
The cases $n=\pm 2$ are treated separately.
We give explicit expressions for the genus zero contribution to the one- and
two-loop correlators as well as for the genus one contribution to the one-loop
correlator and the free energy.
It is shown how one can obtain from these results any multi-loop correlator and
the free energy to any genus and the structure of the higher genera
contributions is described.
Furthermore we describe how the calculation of the higher genera contributions
can be pursued in the scaling limit.

\newpage
\newsection{Introduction}

The $O(n)$ model on a random lattice~\cite{DK88,Kos89} is a matrix model which
regarding its complexity can be placed somewhere in between the one-matrix
model and the two-matrix model.
It is therefore a natural intermediate step if one wants to study the
generalization of 1-matrix model techniques and results to the two- and
eventually the multi-matrix case.
The model is also interesting in its own right having an appealing geometrical
interpretation and a very rich phase structure~\cite{GK89,Kos89,KS92,EZ92}.
In particular when $n=2\cos(\nu\pi)$ with $\nu=l/k$, $0<l<k$ and $l,k\in Z$ the
model has critical points for which the associated scaling behaviour is that
characteristic of 2D gravity interacting with rational conformal matter fields
of the type $(p,q)=(k,(2m+1)k\pm l)$ and with $\nu$ general any central charge
between $c=-\infty$ $(\nu=1)$ and $c=1$ $(\nu=0)$ can be reached.
However, the continuum theories that one obtains from the $O(n)$ model in the
rational case contain only a subset of the operators of the corresponding
minimal models.~\cite{KS92,EZ92}.

In the present paper we will solve the model exactly, i.e.\ without any
assumption of being close to a critical point.
The genus zero contribution to the 1-loop correlator will be calculated solving
the saddle point equation of the model, following the idea of
references~\cite{BIPZ78} and~\cite{EZ92} and the higher genera contributions by
a generalization of the moment technique of reference~\cite{ACKM93}.
As usual this technique will allow us to find from the 1-loop correlator any
multi-loop correlator as well as the free energy.
The parametrization of the coupling constant space of the model in terms of
moment variables reveals that the model possesses the same kind of universality
with respect to the potential as the one-matrix model.
In addition there appears a large degree of universality with respect to $n$.

In the case of the one-matrix model the moment description facilitated the
analysis of the double scaling limit~\cite{dsl}.
For example the result that the continuum 1-matrix model partition function is
a $\tau$-function of the kdV hierarchy~\cite{tau} could easily be understood in
this description~\cite{AK93}, the analysis relying on a representation of the
$\tau$-function as a matrix model, namely the Kontsevich
model~\cite{Kon91,Wit91}, and the moment description of this model~\cite{IZ92}.
The $\tau$-functions of the kdV$_p$ hierarchies with $p>2$ can also be
represented as matrix models, namely as generalized Kontsevich models and
recently the appropriate moment description of these models has been
found~\cite{EYY94,KRI95}.
Hence it should be possible to determine which is the precise relation between
the continuum partition function of the $O(n)$ model, for $n$ rational, and the
$\tau$-functions of the kdV$_p$ hierarchies by comparing the moment description
of the $O(n)$ model with the moment description of the generalized Kontsevich
models.
This requires of course that a d.s.l. relevant version of the moment
description is developed for the $O(n)$ model. A part of our paper will be
devoted to the development of such a description.

We will start by, in section~2, presenting the model and the most important
equations needed for its solution.
Then we will proceed with the exact solution, for $n\in]-2,2[$ in section~3,
and for $n=\pm 2$ in section~4.
Section~5 is devoted to the study of the double scaling limit and section~6
contains our conclusion and a discussion of possible future directions of
investigation.

\newsection{The Model}

In the following we will consider the O(n) model on a random lattice, given by
the partition function
\beq
Z=e^{N^2 F}=\int_{N\times N}dM \prod_{i=1}^{n}dA_i
\exp\left(-N\ltr\left[V(M)+M\sum_{i=1}^nA_i^2\right]\right)
\label{partition}
\eeq
where $M$ and $A_i$, $i=1,\ldots,n$ are hermitian $N\times N$ matrices and
\beq
V(M)=\sum_{j=1}^{\infty}\frac{g_j}{j}M^j.
\eeq
In the language of Feynman diagrams the model describes a gas of $n$ different
types of self-avoiding loops; non-interacting and living on a random
surface\footnote{Strictly speaking, to have this interpretation, we should
include mass terms for the $A$-fields and exclude the term linear in $M$ in our
action.
However, this rearrangement can be obtained by a linear shift of the matrix $M$
and since we will work with a generic potential such a shift can always be
performed in the final result.}~\cite{DK88}.
To begin with $n$ is an integer but by analytical continuation the model can be
defined also for non integer values of $n$.
We will restrict ourselves to the case $|n|\leq 2$ and we will use the
following parametrization
\beq
n=2\cos(\nu \pi),\hspace{0.7cm} 0\leq\nu \leq 1.
\eeq
We note that for $n=0$ the model is identical to the usual hermitian 1-matrix
model.
Furthermore for $n=1$ and a special cubic potential the model describes the
Ising model on a random lattice~\cite{EZ92}.
We shall in particular be concerned with the calculation of the free energy,
$F$, and correlators of the $M$-field of the following type
\beq
W(p_1,\ldots,p_s)=N^{s-2}\left \langle \ltr \frac{1}{p_1-M} \ltr
\frac{1}{p_2-M}
\ldots \ltr \frac{1}{p_s-M}\right \rangle_{conn} \non \\
\eeq
The genus expansion of these objects reads
\beq
F=\sum_{g=0}^{\infty}N^{-2g}F_g,\hspace{1.0cm}
W(p_1,\ldots,p_s)=\sum_{g=0}^{\infty}N^{-2g}W^g(p_1,\ldots,p_s).
\eeq
and we have
\beq
W^g(p_1,\ldots,p_s)
=\frac{d}{dV(p_1)}\ldots \frac{d}{dV(p_s)} F_g,
\hspace{0.7cm}g\geq 1 \mbox{ or }s\geq 2
\label{multiloop}
\eeq
where
\beq
\frac{d}{dV(p)}=-\sum_{j=1}^{\infty} \frac{j}{p^{j+1}} \frac{d}{dg_j}.
\eeq
In the remaining part of this section we shall introduce the tools which will
allow us to determine, for any potential $V(M)$ and any $n\in[-2,2]$,
$W^g(p_1,\ldots,p_s)$ and $F_g$ for (in principle) any $g$ and any $s$.
Eventually it will be convenient to treat separately the cases $n\in ]-2,2[$
and $n=\pm 2$ but here we shall address the aspects which are common to all
values of $n$.

\subsection{The saddle point equation}

The integration over the $A$ matrices in our partition function~\rf{partition}
is gaussian and can directly be carried out.
This leads to the appearance of a 1-matrix integral in which we can diagonalize
the matrices and integrate out the angular degrees of freedom.
By this procedure our partition function (up to a constant) turns into the
following integral over the eigenvalues of the matrix $M$~\cite{Kos89}
\beq
Z\propto \int_{-\infty}^{\infty} \prod_i d\lambda_i e^{-N\sum_i
V(\lambda_i)}\prod_{i,j}
(\lambda_i+\lambda_j)^{-n/2}\prod_{i<j}(\lambda_i-\lambda_j)^2 .
\label{eigenint}
\eeq
In the limit $N\rightarrow \infty$ the eigenvalue configuration is determined
by the saddle point of the integral above~\cite{BIPZ78}.
The corresponding saddle point equation reads~\cite{Kos89}
\beq
V'(\lambda_i)=\frac{2}{N}\sum_{j\neq i}\frac{1}{\l_i-\l_j}-
\frac{n}{N}\sum_{j}\frac{1}{\l_i+\l_j}.
\label{saddle0}
\eeq
As usual this discrete equation can be transformed into a continuous one by
introducing corresponding to the matrix $M$ an eigenvalue density
$\rho(\l)=\frac{1}{N}\sum_i\delta(\l-\l_i)$ which in the limit $N\rightarrow
\infty$ becomes a continuous function~\cite{BIPZ78}.
When one of the eigenvalues approaches the origin, the integral~\rf{partition}
ceases to exist (cf.\ to equation~\rf{saddle0}).
Therefore we will always assume that the eigenvalues are confined to the
positive real axis. More precisely we will consider the situation where the
eigenvalue density has support only on one interval $[a,b]$ on the positive
real axis and is normalized to one, i.e.
\bea
\mbox{supp } \rho(\l)&=&[a,b], \hspace{0.4cm}a>0, \label{condition1}\\
\int_a^b\rho(\l)d\l&=&1. \label{condition2}
\eea
Of course the results obtained in this situation will allow an anlysis of the
case $a\rightarrow 0$.
In terms of the eigenvalue distibution the saddle point equation~\rf{saddle0}
reads~\cite{Kos89}
\beq
V'(\l)=2\int_a^b\hspace{-0.55cm}-\,\,d\mu\frac{\rho(\mu)}{\l-\mu}-n\int_a^bd\mu
\frac{\rho(\mu)}{\l+\mu}.
\eeq
The saddle point equation can also be written in terms of the genus zero
one-loop correlator~\cite{BIPZ78}.
One has
\beq
W^0(p)=\int_a^bd\mu \,\frac{\rho(\mu)}{p-\mu}
\label{Wrho}
\eeq
and the conditions~\rf{condition1} and~\rf{condition2} on $\rho(\l)$ are
equivalent to demanding that $W(p)$ is analytic in the complex plane except
from a cut $[a,b]$ and that
\beq
W(p)\rightarrow \frac{1}{p},\hspace{0.5cm} p\rightarrow \infty.
\label{normalization}
\eeq
The inverse relation to~\rf{Wrho} reads
\beq
\rho(\l)=\frac{1}{2\pi i}\left\{ W^0(\l-i0)-W^0(\l+i0)\right\}
\label{density}
\eeq
and the saddle point equation for $\rho(\l)$ turns into the following equation
for the genus zero contribution the one-loop correlator
\beq
V'(p)=W^0(p+i0)+W^0(p-i0)+nW^0(-p), \hspace{0.7cm} p\in [a,b].
\label{saddle}
\eeq

\subsection{The loop equations}

The loop equations of the model can be derived in various ways~\cite{KS92}.
Here let us use a formulation which exposes very clearly the analogy with the
1-matrix model case.
First we exploit the invariance of the partition function~\rf{partition} under
the following redefinition of the field $M$
\beq
M\rightarrow M+\epsilon\frac{1}{p-M}.
\eeq
Introducing this shift in~\rf{partition} gives rise to the following equation
\beq
\oint_{C_1}\frac{\D\omega}{2\pi i}\frac{ V'(\omega)}{p-\omega}
W(\omega)+
n\chi(p)= (W(p))^2+\frac{1}{N^2}\frac{d}{dV(p)}W(p)
\label{loop1}
\eeq
where
\beq
\chi(p)=\frac{1}{N}\langle\ltr\frac{1}{p-M}A_i^2\rangle
\eeq
and where the contour $C_1$ encloses the cut $[a,b]$ of $W(\omega)$ but not the
point $\omega=p$.
We will use the convention that all contours are oriented counterclockwise.
Next, let us consider the following redefinition of the field $A_i$
\beq
A_i\rightarrow A_i +\epsilon\frac{1}{p-M}A_i \frac{1}{-p-M}.
\eeq
Inserted into~\rf{partition} this shift leads to the following identity
\beq
-\chi(p)-\chi(-p)=W(p)W(-p)+\frac{1}{N^2} \frac{d}{dV(p)} W(-p).
\label{loop2}
\eeq
{}From~\rf{loop1} and~\rf{loop2} we can obtain a closed equation for the 1-loop
correlator of the $M$-field
\bea
\lefteqn{2\oint_{C_1}\frac{\D\omega}{2\pi
i}\frac{V'(\omega)\omega}{p^2-\omega^2}W(\omega) \non } \\
&=&(W(p))^2+(W(-p))^2+nW(p)W(-p) \non \\
&&{}+\frac{1}{N^2}\left\{\frac{d}{dV(p)}W(p)+\frac{d}{dV(-p)}W(-p)+
n\frac{d}{dV(p)}W(-p)\right\}.
\label{loopfinal}
\eea
This equation exhibits a strong similarity with the equation for the 1-loop
correlator of the hermitian 1-matrix model but as opposed to the latter it is
non local.
However, as we shall see in section~\ref{loopeqn2}, \ref{W0pn=-2} and
\ref{W0pn=2} there exists an efficient way to deal with this non-locality.

\newsection{The case $n\in ]-2,2[$}
\subsection{Reformulation of the loop equation \label{loopeqn2} }

With the aim of reformulating~\rf{loopfinal} as a local equation, let us
introduce a function $W_r(p)$ by
\beq
W_r(p)=\frac{2V'(p)-nV'(-p)}{4-n^2}.
\label{Wr}
\eeq
Furthermore, let us corresponding to a function or an operator $h(p)$ define
$h_{\pm}(p)$ by
\beq
h_+(p)=\frac{e^{+i\nu \pi/2}h(p)+e^{-i\nu\pi /2}h(-p)}{2\sin(\nu \pi)},
\hspace{0.7cm}h_-(p)=h_+(-p).
\label{rotate}
\eeq
Inversely we then have
\beq
h(p)=-i\left(e^{i\nu\pi/2} h_+(p)-e^{-i\nu \pi /2} h_-(p)\right).
\label{invrot}
\eeq
Introducing the transformation~\rf{rotate} into the loop
equation~\rf{loopfinal} one gets
\bea
\lefteqn{\oint_{C_2} \frac{\D\omega}{2 \pi i} \frac{\om}{p^2-\om^2}
\left\{W_{r+}(\omega)
W_-(\omega)
+W_{r-}(\omega)W_+(\omega)\right\}\hspace{1.0cm} \non} \\
&&\hspace{4.0cm}=
W_+(p)W_-(p)+\frac{1}{N^2}\frac{d}{dV_+(p)}W_-(p)
\label{looppm}
\eea
where the contour, $C_2$, now encircles $[a,b]$ as well as $[-b,-a]$ but not
the point $\omega=p$ and where $d/dV_+(p)$ is shorthand notation for
$\left(d/dV(p)\right)_+$.
Introducing two linear operators $\hat{K}_+$ and $\hat{K}_-$ by
\beq
\hat{K}_+f(p)=\oint_{C_2} \frac{\D\omega}{2 \pi i}
\frac{\omega W_{r+}(\omega)}{p^2-\omega^2}f(\omega),\hspace{0.7cm}
\hat{K}_-f(p)=\oint_{C_2} \frac{\D\omega}{2 \pi i}
 \frac{\omega W_{r-}(\omega)}{p^2-\omega^2}f(\omega)
\eeq
and inserting the genus expansion of the correlators in~\rf{looppm} we find
\bea
\lefteqn{
\left\{\hat{K}_+-W_+^0(p)\right\}W_-^g(p)+
\left\{\hat{K}_--W_-^0(p)\right\}W_+^g(p) \non} \\
&&\hspace{1.5cm}=\sum_{g'=1}^{g-1}W_+^{g'}(p)W_-^{g-g'}(p)+
\frac{d}{dV_+(p)}W_-^{g-1}(p),\hspace{0.7cm}g\geq 1.
\label{loopgen}
\eea
The similarity with the corresponding equation of the hermitian 1-matrix model
appearing in reference~\cite{ACKM93} is striking and we will later show how the
strategy of reference~\cite{ACKM93} for solving the loop equation genus by
genus can be generalized to the present case.
Of course an iterative procedure for solving~\rf{loopgen} requires the
knowledge of $W^0(p)$.
In the next section we will show how one can write down a closed expression for
this correlator, i.e.\  an expression which is valid for any potential $V(M)$
and any $n\in ]-2,2[$.

\subsection{$W^0(p)$ in terms of an auxiliary function $G(p)$
\label{genus0sol} }

To determine the 1-loop correlator at genus zero we follow the idea of
reference~\cite{EZ92}.
As mentioned earlier we restrict ourselves to the one-cut situation.
Our starting point will be the saddle point equation \rf{saddle} which together
with the boundary condition~\rf{normalization} determines uniquely $W^0(p)$.
Let us split $W(p)$ in a regular part $W_r(p)$ and a singular part $W_s(p)$
\beq
W(p)=W_r(p)-W_s(p).
\label{Wrs}
\eeq
{}From~\rf{saddle} it follows that $W_r(p)$ is given by~\rf{Wr} while
$W_s^0(p)$ obeys the homogeneous saddle point equation and the boundary
equation
\beq
W_s^0(p)\sim W_r(p)-\frac{1}{p},
\hspace{0.7cm} p\rightarrow \infty.
\label{boundary}
\eeq
In the language of the rotated functions $W_{\pm}(p)$ we have the following
situation
\beq
W_{\pm}^0(p)=W_{r\pm}(p)-W_{s\pm}^0(p)
\eeq
with $W_{r+}(p)$ being given by
\beq
W_{r+}(p)=i\;\frac{e^{-i\nu \pi/2}V'(p)-e^{i\nu\pi /2}V'(-p)}{4-n^2}
\eeq
and with $W^0_{s\pm}(p)$ obeying the equations~\cite{EZ92}
\beq
W^0_{s\pm}(p-i0)=-e^{\pm i \nu \pi} W^0_{s\mp}(p+i0).
\label{newsaddle}
\eeq
The boundary equation~\rf{boundary} translates into
\beq
W^0_{s+}(p)\sim W_{r+}(p)-\frac{i}{2\cos(\nu \pi /2)}\;\frac{1}{p}.
\label{bound+}
\eeq
In order to obtain a closed expression for $W^0_{s\pm}(p)$ we introduce an
auxiliary function $G(p)$ with the following properties
\begin{enumerate}
\item
$G_{\pm}(p)$ fulfill the equations~\rf{newsaddle}.
\item
$G(p)$ is analytic in the complex plane except from a cut $[a,b]$ and behaves
as $(p-a)^{-1/2}(p-b)^{-1/2}$ in the vicinity of $a$ and $b$.
\item
$G_{\pm}(p) \sim \pm \frac{i}{p},\hspace{0.7cm} p\rightarrow \infty$.
\end{enumerate}
\label{Gcond}
In section~\ref{Gexpl} we will show that these requirements are enough to fix
$G(p)$ uniquely.
For the moment let us note that 1--3\ imply that the function
$R(p)=G_+(p)G_-(p)$ is even, behaves as $1/p^2$ as $p\rightarrow \infty$ and
can have no singularities except for single poles for $p=\pm a, \pm b$, i.e.
\beq
R(p)=G_+(p)G_-(p)=\frac{(p^2-e^2)}{(p^2-a^2)(p^2-b^2)}.
\label{G+-}
\eeq
We will choose the convention that $+e$ is a root of $G_+(p)$ while $-e$ is a
root of $G_-(p)$.
We will later write down an equation which determines $e$ in terms of $a$ and
$b$.
Now if we write a generic solution of~\rf{newsaddle}, $S_{\pm}$, as
\beq
S_{\pm}(p)=\overline{S}_{\pm}(p)G_{\pm}(p)
\label{WtildeG}
\eeq
we have
\beq
\overline{S}_{\pm}(p-i0)=\overline{S}_{\mp}(p+i0)
\eeq
which means that the even function $\overline{S}_+(p)+\overline{S}_{-}(p)$ is a
regular function while the odd function $\overline{S}_+(p)-\overline{S}_-(p)$
has a square root branch cut $[a,b]$.
Hence we have
\beq
\overline{S}_+(p)=A(p^2)+pB(p^2)\overline{\sqrt{p}},\hspace{0.7cm}
\overline{\sqrt{p}}=\sqrt{(p^2-a^2)(p^2-b^2)}
\eeq
with $A(p^2)$ and $B(p^2)$ regular but not necessarily entire functions.
Since $+e$ is a root of $G_+(p)$, $A(p^2)$ and $B(p^2)$ may have a pole for
$p=e$ without $S_{+}(p)$ becoming singular there provided the accompanying pole
for $p=-e$ is cancelled (cf.\ equations~\rf{G+-} and~\rf{WtildeG}).
Since we are not particularly interested in solutions which vanish for $p=e$ a
more convenient parametrization is
\beq
S_{+}(p)=A(p^2)G_+(p) +pB(p^2)g_+(p)G_+(p)
\label{para1}
\eeq
where
\beq
g_+(p)=\frac{\overline{\sqrt{p}}+\frac{p}{e}\overline{\sqrt{e}}}{p^2-e^2}
\eeq
and where $A$ and $B$ are again regular but not necessarily entire functions.
We can also write
\beq
S(p)=A(p^2)G(p)+pB(p^2)\tilde{G}(p)
\label{S(p)}
\eeq
where
\beq
\tilde{G}(p)=-\left(e^{i\nu \pi/2} g_+(p)G_+(p)+e^{-i\nu\pi/2}g_-(p)G_-(p)
\right)
\label{Gtilde}
\eeq
We draw the attention of the reader to the equation~\rf{S(p)}.
This equation will play a key role throughout the paper.
It says that any solution of the saddle point equation~\rf{newsaddle} can be
parametrized in terms of the two functions $G(p)$ and $\tilde{G}(p)$.
In particular one has that any such solution can be parametrized in terms  of
any two other independent solutions.
A study of the analyticity properties of $\tilde{G}(p)$ reveals an interesting
symmetry of the model.
Let us for a moment write the function $G(p)$ as $G^{\nu}(p)$ where the index
$\nu$ is the parameter which enters the relation $n=2\cos(\nu \pi)$.
Then we have
\beq
\tilde{G}^{\nu}(p)=G^{(1-\nu)}(p).
\label{GGtilde}
\eeq
This follows from the fact that $\tilde{G}(p)$ is a solution of the saddle
point equation~\rf{newsaddle} with $\nu$ being replaced by $1-\nu$.
Furthermore from~\rf{GGtilde} it follows that the parameter $e=e_{\nu}$
entering the relation~\rf{G+-} for $G^{\nu}(p)$ is related to the corresponding
parameter, $e_{1-\nu}$, for $G^{1-\nu}(p)$ by~\footnote{The choice of sign in
this relation is a rather technical point. It relies on the
expression~\rf{Gelliptbound} in the next section. We note, however, that all
physical quantities depend only on $e^2$.}
\beq
e_{1-\nu}=-\frac{ab}{e_{\nu}}
\label{eetilde}
\eeq
Let us now specialize to the 1-loop correlator.
Since we want this function to be finite in the limits $p\rightarrow a,b$ we
choose in this case a slightly different but equivalent parametrization, namely
\beq
W_{s+}^0(p)=\overline{\sqrt{p}}\left({\cal A}(p^2)g_+(p)G_+(p)+
p{\cal B}(p^2)G_+(p)\right).\label{gensol}
\eeq
Due to the assumptions concerning the analyticity properties of the 1-loop
correlator ${\cal A}(p^2)$ and ${\cal B}(p^2)$ must here be entire functions
and from the boundary condition~\rf{bound+} it follows that they are
necessarily polynomials.
Using the relation~\rf{G+-} one easily concludes that ${\cal A}(p^2)$ and
${\cal B}(p^2)$ can be expressed in the following way
\bea
{\cal A}(p^2)&=&\frac{1}{2}\left(G_-(p)W_{s+}^0(p)+G_+(p)W_{s-}^0(p)\right), \\
p{\cal B}(p^2)&=&\frac{1}{2}\left(g_-(p)G_-(p)W_{s+}^0(p)-
g_+(p)G_+(p)W_{s-}^0(p)\right).
\eea
The fact that ${\cal A}(p^2)$ and ${\cal B}(p^2)$ are polynomials and that
$W_{s+}^0(p)\sim W_{r+}(p)+{\cal O}(1/p)$ allows one to conclude
\bea
{\cal A}(p^2)&=&Even\mbox{ } polynomial\mbox{ } part\mbox{ } of \mbox{ }
W_{r-}(p)G_{+}(p) \non\\
&=&\oint_{\infty}\frac{\D\omega}{2\pi i}
\frac{\omega W_{r-}(\omega)}{p^2-\omega^2}G_+(\omega) \\
p{\cal B}(p^2)&=& - Odd\mbox{ } polynomial \mbox{ }part\mbox { } of\mbox{ }
W_{r-}(p)g_+(p)G_+(p) \non \\
&=& -p\;\oint_{\infty} \frac{\D\omega}{2\pi i}
\frac{W_{r-}(\omega)}{p^2-\omega^2}
g_+(p)G_+(\omega)
\eea
where $\oint_{\infty}$ means integration along a contour which encircles
$\infty$.
In total one can write $W_+^0(p)$ as the following contour integral
\beq
W_+^0(p)=\overline{\sqrt{p}}G_+(p)\oint_{C_2}
\frac{\D\omega}{2\pi i} \frac{W_{r-}(\omega)}{p^2-\omega^2}G_+(\om)
\left\{ \om g_+(p)-pg_+(\om)\right\}
\label{W0+}
\eeq
where the contour $C_2$ encircles the cuts $[a,b]$ and $[-b,-a]$ of
$G_+(\omega)$ but not the point $\omega=p$.
The points $a$ and $b$ are determined by the following two equations
\bea
&&\oint_{C_2}\frac{\D\omega}{2\pi i}W_{r-}(\omega)g_+(\om)G_+(\omega) =0,
\label{B1+}  \\
&&\oint_{C_2}\frac{\D\omega}{2\pi i} W_{r-}(\omega)G_+(\omega) \omega
= \frac{-1}{2\cos(\nu\pi/2)}
\label{B2+}
\eea
which follow from the boundary condition~\rf{normalization}.
We note that by using the analyticity properties of the various functions
entering the integrand~\rf{W0+} one can replace
$\oint_{C_2}\frac{\D\omega}{2\pi i}W_{r-}(\omega) \left\{\ldots\right\}$ by
$-\frac{e^{-i\nu \pi/2}}{2sin(\nu \pi)} \oint_{C_1}\frac{\D\omega}{2\pi
i}V'(\omega)\left\{\ldots\right\}$ in the expressions~\rf{W0+}--\rf{B2+}.
It is a matter of taste which expression one prefers to work with.
The former reflects more clearly the structure of the loop equation while the
latter expression ressembles more the one of the hermitian 1-matrix model.

\subsection{Determination of the auxiliary function \label{Gexpl} }
\subsubsection{General case}

One can derive a differential equation for $G_+(p)$.
To do so one first observes that the function $\frac{\partial}{\partial
p}\left(\overline{\sqrt{p}}G_+(p) \right)$ will fulfill the
equation~\rf{newsaddle} and hence have a parametrization of the
type~\rf{para1}.
When supplemented by the boundary condition for $G_+(p)$ this observation
allows one to conclude
\beq
\frac{\partial}{\partial p}\left(\overline{\sqrt{p}}G_+(p)\right)=
\left(\alpha-\frac{\overline{\sqrt{e}}}{e}+pg_+(p)
\right)G_+(p)
\label{diffp}
\eeq
where $\alpha$ is some yet not determined constant which has the following role
\beq
G_+(p)=\frac{i}{p}\left(1-\frac{\alpha}{p}+{\cal
O}\left(\frac{1}{p^2}\right)\right),\hspace{0.7cm} p\rightarrow \infty.
\label{alpha}
\eeq
For given $e$ and $\alpha$ the equation~\rf{diffp} determines $G_+(p)$
uniquely.
It is easy to see that $G_+(p)$ given by the following elliptic integral is the
unique solution we seek
\beq
\log\left(\frac{\overline{\sqrt{p}}\;G_+(p)}{\sqrt{p^2-e^2}}\right)=
\int_0^p\frac{dx}{\overline{\sqrt{x}}}\left(
\frac{e\overline{\sqrt{e}}}{x^2-e^2}+\alpha\right)
\label{Gelliptic}
\eeq
provided the following two equations hold
\beq
\int_b^{\infty}\frac{dx}{\overline{\sqrt{x}}}
\left(\frac{e\overline{\sqrt{e}}}{x^2-e^2}+\alpha\right)=\frac{i(1-\nu)\pi}{2},
\hspace{0.7cm}
\int_0^{a}\frac{dx}{\overline{\sqrt{x}}}
\left(\frac{e\overline{\sqrt{e}}}{x^2-e^2}+\alpha\right)=\frac{i\nu\pi}{2}.
\label{Gelliptbound}
\eeq
These equations ensure that $G_+(p)$ has the correct asymptotic behaviour as
$p\rightarrow \infty$ and that $G_+(a+i0)=-e^{i\nu\pi/2}G_-(a-i0)$ (cf.\ to
equation~\rf{newsaddle} ).
Together they determine the unknowns $e$ and $\alpha$.
In particular it can be shown that $e$ must necessarily lie on the positive
imaginary axis and behave as $a^{\nu}$ when $a\rightarrow 0$.
One can derive another set of equations which determines these two quantities
and which will be of importance for the analysis in the following sections.
Using the same strategy as for the derivation of~\rf{diffp} one finds the
following expression for the derivative of $G_+(p)$ with respect to $a^2$
\beq
\frac{\partial}{\partial a^2}G_+(p)=\frac{1}{p^2-a^2}\left(
\lambda_a p g_+(p) +\frac{1}{2}(1-\rho_a)\right)G_+(p)
\label{diffa}
\eeq
where
\beq
\lambda_a=-\frac{1}{2}\frac{\partial e^2}{\partial
a^2}\frac{e^2-a^2}{e\overline{\sqrt{e}}},
\hspace{0.7cm}\rho_a=\frac{a^2}{e^2}\frac{\partial e^2}{\partial a^2}
\label{lambdarho}
\eeq
Now comparing the expressions for $\frac{\partial}{\partial
a^2}\frac{\partial}{\partial p}G_+(p)$ and $\frac{\partial}{\partial
p}\frac{\partial}{\partial a^2}G_+(p)$ that one obtains from~\rf{diffp}
and~\rf{diffa} respectively, one finds the following relation between $a$, $b$,
$e$ and $\alpha$
\bea
\alpha&=&-{e\overline{\sqrt{e}}\over a^2-e^2}+\frac{\partial e^2}{\partial
a^2}\frac{a^2(b^2-a^2)}
{e\overline{\sqrt{e}}}, \\
\frac{\partial \alpha}{\partial a^2}&=&-\lambda_a. \label{dada}
\eea
In particular these two equations allow one to write down a second order
differential equation for $e(a,b)$.
We shall refrain from doing so since we have not been able to extract any
further information about the model from the resulting equation.
Let us for later convenience note that we have also the relation
\beq
\frac{\partial}{\partial p}\left(\overline{\sqrt{p}}g_+(p)G_+(p)\right)
=\left(\alpha g_+(p)+p\right)G_+(p)
\label{diffp2}
\eeq
as well as
\beq
\frac{\partial}{\partial a^2}\left(pG_+(p)\right)=
\frac{e\overline{\sqrt{e}}}{e^2-b^2}
\frac{\partial}{\partial a^2}\left(g_+(p)G_+(p)\right)+\lambda_a g_+(p)G_+(p)
\label{diffa2}
\eeq
Here~\rf{diffp2} follows immediately from~\rf{GGtilde} by noting that for
$g_+(p)G_+(p)$ the parameter $\alpha$ entering~\rf{alpha} is replaced by
$\tilde{\alpha}$,
\beq
-\tilde{\alpha}=\alpha-\frac{\overline{\sqrt{e}}}{e}
\eeq
and to derive the relation~\rf{diffa2} one makes use of the fact that any two
solutions of the saddle point equation can be parametrized in terms of any two
other independent solutions.
The detailed nature of the parametrization follows from the analyticity
properties and the asymptotic behaviour of the functions involved.
Needles to say that relations similar to~\rf{diffa} and \rf{diffa2} concerning
the differentiation with respect to $b^2$ follow from these by the
interchangements $a \leftrightarrow b$ and that $e$ and $\alpha$ depend on $a$
and $b$ in a symmetrical manner.
As we will show in the next section when $\nu$ is rational $G_+(p)$ can be
further explicited.

\subsubsection{Rational case}

Let us parametrize $\nu$ in the following way
\beq
\nu=\frac{l}{q}, \hspace{0.8cm}0<l<q,\hspace{0.4cm} l,q\in Z_+
\eeq
and let us following reference~\cite{EZ92} introduce the function
\beq
 T(p)=\frac{1}{2}\left\{ (G_+(p))^q + (-1)^{q+l} (G_-(p))^q \right\}.
\label{Tp}
\eeq
{}From the requirements 1--3 on $G(p)$ it follows that $T(p)$ is a rational
function with poles at $\pm a$ and $\pm b$ of order $[q/2]$ (the integer part
of $q/2$).
Furthermore from~\rf{G+-} and~\rf{Tp} it follows that $\left(G_+(p)\right)^q$
can be expressed via the two rational functions $T(p)$ and $R(p)$ in the
following way
\beq
\left(G_+(p)\right)^q= T(p) - \sqrt{ T(p)^2-(-1)^{l+q} R(p)^q }
\label{G+q}
\eeq
where the negative sign in front of the square root ensures the correct
asymptotic behaviour of $G_+(p)$ as $p\rightarrow \infty$.
Now the requirement that $G(p)$ must be analytic in the complex plane except
from a cut $[a,b]$ implies that the the square root term above can have
singularities only at $a$ and  $b$ and therefore must decompose as
$\tilde{T}(p)\overline{\sqrt{p}}$ with $\tilde{T}(p)$ another rational
function.
Hence we can parametrize $G_+(p)$ in the following way
\beq
G_+(p)=\frac{i}{\overline{\sqrt{p}}}\left\{\, (p^2-e^2)\,\left[\, A(p)g_-(p) +
B(p)\,\right]\, \right\}^{1/q}
\label{G+rational}
\eeq
where $A(p)$ and $B(p)$ are polynomials of degree less than or equal to $q-2$
and where we have made use of the function $g_-(p)$ in order to obtain the
property $G_+(e)=0$ assumed earlier.
Noting that in the  relation~\rf{G+q} both $T(p)$ and the function appearing
under the square root, for given $l$ and $q$, are functions of a definite
parity one finds that the same must be true for $A(p)$ and $B(p)$.
More precisely
\beq
 A(-p)=(-1)^{l+1} A(p), \hspace{0.7cm} \hs B(-p)=(-1)^l B(p).
\label{lparity}
\eeq
To determine the polynomials $A(p)$ and $B(p)$ as well as the parameter $e$ it
suffices to evoke the relation~\rf{G+-} which implies
\beq
 (-1)^{q+l}\,(p^2-e^2)^{q-1} \, = \, (p^2-e^2) B^2(p) -
\left(p^2-\frac{a^2b^2}{e^2}\right) A^2(p) \, -2
\frac{\overline{\sqrt{e}}}{e}p A(p) B(p).
\label{alg}
\eeq
and where we note that the number of equations exactly matches the number of
unknowns.
However, this set of algebraic equations may have many different solutions and
we must add some boundary condition to select the correct one.
Let us note that equations~\rf{G+rational},~\rf{lparity} and~\rf{alg} do not
depend on $l$ but only on its parity.
We claim that the different solutions of equation~\rf{alg} correspond to
different values of $l$.
For a given $l$ the correct solution can be identified for instance by its
asymptotic behaviour in the $a\to 0$ limit.
As mentioned in the previous section $e(a,b)$ always lies on the positive
imaginary axis and in the limit $a\rightarrow 0$, it behaves as $a^{\nu}$.
More precisely as we shall see in section~\ref{Oncrit} one has
\beq
e(a,b) \mathop{\sim}_{a\to 0} 2ib \left({a\over 4b}\right)^{l/q}
\label{ecrit}
\eeq
and this is the criterion which allows us to pick out a unique solution of
equation~\rf{alg}.
One might prefer evaluating the logarithmic derivative $\rho_a$ introduced
in~\rf{lambdarho} which must behave as
\beq
\hs \rho_a \mathop{\sim}_{a\to 0} \nu={l\over q}.
\label{rhoacritlq}
\eeq
Let us close this section by considering some explicit examples.
In each case the function $G(p)$ is determined by the
equations~\rf{G+rational},~\rf{lparity},~\rf{alg} and~\rf{ecrit}
or~\rf{rhoacritlq}.
 \newline
\newline
{\bf The case l=1, q=2, i.e. n=0:}
Here equations~\rf{lparity},~\rf{alg} and the condition that the degree of $A$
and $B$ is less than $q-2$ imply that
\beq
A(p)=1,\hspace{0.7cm} B(p)=0 \hs \hbox{and} \hs e^2={a^2 b^2\over e^2} \hs
\hbox{i.e.}\hs e=+i\sqrt{ab}.
\eeq
The expression~\rf{G+rational} for $G_+(p)$ hence reads
\beq
G_+(p)=\frac{i}{\overline{\sqrt{p}}}\left\{\overline{\sqrt{p}}
-p\frac{\overline{\sqrt{e}}}{e}\right\}^{1/2}=\frac{i}{\overline{\sqrt{p}}}\left\{\overline{\sqrt{p}}
-i p (b-a)\right\}^{1/2}.
\label{G+n0}
\eeq
After performing the transformation~\rf{invrot} one finds the familiar form of
the solution of the saddle point equation of the hermitian 1-matrix model with
behaviour $G(p)\rightarrow \frac{\sqrt{2}}{p}$, $p\rightarrow \infty$
\beq
G(p)=\frac{\sqrt{2}}{{\sqrt{(p-a)(p-b)}}}.
\eeq
Furthermore one easily  verifies that with the expression~\rf{G+n0} for
$G_+(p)$ the formulas in section~\rf{genus0sol} correctly reproduce the usual
contour integral representation of the solution of the 1-matrix model.
\newline
\newline
{\bf The case l=1, q=3, i.e. n=1:} Let us emphasize that this
set of models contains the Ising model on a random lattice as a special case.
Since the polynomials $A(p)$ and $B(p)$ are of degree less than or equal to
$q-2=1$ and obey the parity condition~\rf{lparity} we write them in the
following way:
\beq
A(p)=c,\hspace{0.7cm} B(p)=p
\eeq
The constant $c$ and the parameter $e$ are determined by~\rf{alg}.
For $c$ one finds
\beq
c=\frac{-2e\overline{\sqrt{e}}}{e^2-a^2b^2/e^2}
\eeq
while $e$ is given by
\beq
e^2=-\epsilon a b,\hspace{0.7cm}\eps^4-6 \eps^2-4\eps
\left(\frac{a}{b}+\frac{b}{a}\right)-3=0
\eeq
According to~\rf{ecrit} and~\rf{rhoacritlq} we have to choose the branch of the
solution of this fourth degree equation for which $\epsilon>0$ and
$\rho_a\rightarrow 1/3$ when $a\rightarrow 0$, i.e.\ when $\epsilon\rightarrow
\infty$.
For the solution which matches these criteria one has
\beq
\rho_a={1\over 2}\left( 1-{1\over 3}\sqrt{\frac{\epsilon^2-9}{\epsilon^2-1}}
\right)
\eeq
where it is understood that the positive square root should be taken.

\subsection{The two-loop correlator at genus zero \label{twoloop} }

One way to calculate the two-loop correlator is to use the directly the recipe
\beq
W(p,q)=\frac{d}{d V(p)}W(q). \label{W++pq}
\eeq
However, there exists a less work demanding method.
The two-loop correlator at genus zero must satisfy the following equation
\beq
 W^0(p+i0,q)+W^0(p-i0,q)+n W^0(-p,q)=-\frac{1}{(p-q)^2},\hspace{0.7cm}
p\in[a,b]
\label{saddle2loop}
\eeq
which appears when one applies the loop insertion operator $d/dV(q)$ to the
saddle point equation~\rf{saddle}.
This is an equation of the same type as~\rf{saddle}.
One can split the two-loop correlator in a regular and a singular part.
The regular part is easily found and coincides with what one finds by acting
with the loop insertion operator on the regular part of the one-loop
correlator.
The singular part fulfills the homogeneous version of the
equation~\rf{saddle2loop}.
To solve this equation it is convenient to perform a rotation like~\rf{rotate}
for each of the two variables of $W(p,q)$ so that one has
\beq
W(p,q)=-e^{i\nu\pi}W_{++}(p,q)+W_{+-}(p,q)+W_{-+}(p,q)-e^{-i\nu\pi}W_{--}(p,q).
\eeq
with
\beq
W_{+-}(p,q)=W_{--}(-p,q)=W_{-+}(-p,-q)=W_{++}(p,-q).
\eeq
We note that
\beq
W_{i,j}(p,q)=\frac{d}{dV_i(p)}W_j(q),\hspace{0.7cm} i,j\in \{+,-\}.
\eeq
Now the singular part of $W_{++}^0(p,q)$ fulfills an equation similar to
\rf{newsaddle} in each of the variables and a parametrization of the most
general solution can be written down using the functions $G_+(p)$ and
$g_+(p)G_+(p)$ introduced in section~\rf{genus0sol}.
The following requirements on $W(p,q)$ single out a unique solution.
\begin{itemize}
\item $W(p,q)=\frac{d}{dV(p)} \frac{d}{dV(q)} F$ must be symmetrical in $p$ and
$q$  and  regular when $p=q$
\item $W^0(p,q)$ can have a singularity of the form $((p-a)(p-b))^{-1/2}$ but
no additional poles at $a$ or $b$ since $W^0(p)$ has only a singularity of the
type $\left((p-a)(p-b)\right)^{1/2}.$
\item $W(p,q)$ has the following asymptotic behaviour
\beq
W(p,q) \sim O(1/p^2),\hspace{0.7cm} p \rightarrow \infty.
\eeq
\end{itemize}
The unique solution reads
\bea
W_{++}^0(p,q)&=&\frac{1}{4-n^2}\left\{G_+(p)G_+(q)\left[-1-\alpha\frac{p
g_+(p)-
q g_+(q)}{p^2-q^2} \right.\right.\non \\
&& \left.\left.+\frac{\left(p g_+(p)-q g_+(q)\right)
\left(p\overline{\sqrt{p}}-q\overline{\sqrt{q}}\right)}{(p^2-q^2)^2}\right]-
\frac{1}{(p+q)^2}\right\}
\label{W++}
\eea
%where
%\beq
%f_+(p)=p g_+(p)-\frac{\overline{\sqrt{e}}}{e}=\frac{\overline{p\sqrt{p}}+
%%e\overline{\sqrt{e}}}{p^2-e^2}
%\eeq
We see that the result does not show any explicit dependence of the matrix
model potential.
Hence the universality of the two-loop function observed for the 1-matrix
model~\cite{AJM90,BZ93,Eyn93} extends to the $O(n)$ model on a random lattice.
In addition there is a large degree of universality with respect to $n$.
(We remind the reader that the result above is valid for any $n$, but that
different values of $n$ give rise to different functions $G_+(p)$.)

In accordance with the fact that $W_+^0(p)$ depends on the potential $V(p)$
only via $W_{r-}(p)$ and that
\beq
\frac{d}{dV_+(q)}W_{r-}(p)=\frac{1}{4-n^2}\frac{\partial}{\partial
q}\left(\frac{1}{p-q}\right)\label{dW-dV+}
\eeq
we find that the two-loop correlator can be written as a total derivative
\beq
W_{++}^0(p,q)=\frac{1}{4-n^2}\frac{\partial}{\partial q}\left\{
-\frac{1}{p+q}+G_+(p)G_+(q)\overline{\sqrt{q}}\;\frac{p g_+(p)-q
g_+(q)}{p^2-q^2}
\right\}.
\eeq
To proceed with the solution of the loop equation we need to know
$W_{+-}^0(p,p)$.
To determine this quantity we must analyse carefully the limit $p\rightarrow q$
of $W_{++}^0(p,-q)$ (which is a rather time consuming task).
The outcome of the analysis is
\beq
W_{+-}^0(p,p)=\frac{1}{4-n^2}\left\{\frac{1}{2}
\frac{e^2-\alpha^2}{(p^2-a^2)(p^2-b^2)}+
\frac{1}{4}(a^2-b^2)^2\frac{p^2}{(p^2-a^2)^2(p^2-b^2)^2}\right\}.
\label{W+-pp}
\eeq
We draw the attention of the reader to the fact that $W_{+-}(p,p)$ is a
rational even function with poles at $p=\pm a$ and $p=\pm b$.
This will be of importance for the following.

\subsection{The one-loop correlator at genus 1 \label{1loop} }
\subsubsection{The structure of the 1-loop correlator}

For genus 1 the loop equation reduces to
\beq
 \frac{\hat{K}}{4-n^2} W^{(1)}(p)=W^0_{+-}(p,p)
\eeq
where $\frac{\hat{K}}{4-n^2}$ is the linear operator entering the left hand
side of the loop equation~\rf{loopgen}, i.e.\
\beq
\frac{\hat{K}}{4-n^2} f(p)= \left\{\hat{K}_+-W_+^0(p)\right\}f_-(p)+
\left\{\hat{K}_--W_-^0(p)\right\}f_+(p).
\eeq
Let us note for later convenience that using the decomposition~\rf{Wrs} we can
write the action of the operator $\hat{K}$ on a function $f(p)$ as
\bea
\hat{K} f(p) & = & (4-n^2)\oint_{{\cal C}_2}\frac{d\om}{2\pi i}  {\omega
\over p^2-\omega^2} \left\{ W_{s+}(\omega)f_-(\omega)+
W_{s-}(\omega)f_+(\omega) \right\} \label{Koint+-} \non \\
 & = &2\,(4-n^2)\,\cdot\,\,{even\,\: fractional\,\:part\,\: of }\,\,\,
W_{s+}(p)f_-(p).  \label{fractional}
\eea
Noticing the structure of expression~\rf{W+-pp} for $W^0_{+-}(p,p)$ and bearing
in mind the strategy for calculating higher genera contributions in the case of
the hermitian 1-matrix model~\cite{ACKM93} we will seek to express $W^1(p)$ in
the following way
\beq
W^1(p)=A_1^{(2)} \chi_a^{(2)}(p)+B_1^{(2)}
\chi_b^{(2)}(p)+A_1^{(1)}\chi_a^{(1)}(p)+B_1^{(1)}\chi_{b}^{(1)}(p)
\label{W1p}
\eeq
the idea being that the $\chi$-functions should allow us to invert the operator
$\hat{K}$, i.e.\
\beq
\hat{K}\chi_a^{(m)}(p)=\frac{1}{(p^2-a^2)^m},\hspace{0.7cm}
\hat{K}\chi_b^{(m)}(p)=\frac{1}{(p^2-b^2)^m},\hspace{0.7cm}
\label{eigenfunction}
\eeq
We will require that the $\chi$-functions have the following asymptotic
behaviour
\beq
\chi_{a}^{(m)}(p),\chi_b^{(m)}(p) \mathop{\sim}_{p\to \infty} {\cal
O}\left(\frac{1}{p^2}\right) \label{boundchi}
\eeq
where the possibility of a $1/p$ term has been excluded in order to ensure that
the relation~\rf{normalization} remains true.
The coefficients $A_1^{(i)}$ and $B_1^{(i)}$, $i=1,2$ follow from the
decomposition of $W^0_{+-}(p,p)$ into fractions of the type $(p^2-a^2)^{-m}$
and $(p^2-b^2)^{-m}$, $m=1,2$ and read
\bea
A_1^{(1)} =\frac{1}{2}\frac{1}{a^2-b^2}\left(
e^2-\alpha^2-\frac{1}{2}(a^2+b^2)\right), && A_1^{(2)} =
\frac{1}{4}a^2,\label{A}\\
B_1^{(1)}
=\frac{1}{2}\frac{1}{b^2-a^2}\left(e^2-\alpha^2-\frac{1}{2}(b^2+a^2)\right), &&
B_1^{(2)}=\frac{1}{4}b^2.  \label{B}
\eea

\subsubsection{Determination of the $\chi$-functions \label{chidet} }

Since the analyticity structure of the $\chi$-functions should be compatible
with that of the 1-loop correlator it is natural to try to construct these
functions using as starting point the functions $G(p)$ and $p\tilde{G}(p)$
(cf.\ to equation~\rf{S(p)}).
{}From~\rf{fractional} it follows that
\beq
\hat{K}G(p)=\hat{K}p\tilde{G}(p)=0.
\eeq
Next, let us consider the following functions
\beq
\phi_a^{(k)}(p)=\frac{G(p)}{(p^2-a^2)^k},\hspace{0.7cm}
\tilde{\phi}_a^{(k)}(p)=\frac{p\tilde{G}(p)}{(p^2-a^2)^k}.
\eeq
Applying the operator $\hat{K}$ to these functions one finds
\beq
\hat{K}\phi_a^{(k)}=\sum_{l=1}^k\frac{m_{a,l}^{(k)}}{(p^2-a^2)^l},\hspace{0.7cm}
\hat{K}\tilde{\phi}_a^{(k)}=\sum_{l=1}^k\frac{\tilde{m}_{a,l}^{(k)}}
{(p^2-a^2)^l}
\eeq
where $\{m_{a,l}^{(k)}\}$ and $\{\tilde{m}_{a,l}^{(k)}\}$ are some constants.
This means that from either of the two series of functions $\phi_a^{(k)}(p)$
and $\tilde{\phi}_a^{(k)}(p)$ we can construct functions $\chi_a^{(m)}(p)$
obeying~\rf{eigenfunction} and~\rf{boundchi}.
(We remind the reader that $G(p),\tilde{G}(p)\sim {\cal O}(1/p)$, $p\rightarrow
\infty$.)
However, neither of the two series alone can serve as building blocks for
$\chi_a^{(m)}(p)$ since all the functions $\phi_a^{(k)}(p)$ and
$\tilde{\phi}_a^{(k)}(p)$ have poles at $p=-a$ which contradicts the assumption
concerning the analyticity structure of $W(p)$.
We are henced forced to take linear combinations of $\phi$'s and
$\tilde{\phi}$'s to kill these unwanted poles.
One type of such linear combinations with correct analyticity properties leaps
to the eye; the functions $\left(\frac{\partial}{\partial a^2}\right)^k G(p)$
(cf.\ to equation~\rf{diffa}).
Let us try to build $\chi_a^{(k)}(p)$ from such functions.
The expression for $\frac{\partial G(p)}{\partial a^2}$ appeared in
equation~\rf{diffa} and for $\frac{\partial^2 G(p)}{\partial (a^2)^2}$ we find
using~\rf{diffa} and~\rf{diffa2}
\beq
\left(\frac{\partial}{\partial a^2}\right)^2G(p)
=\frac{3}{2}\frac{1}{p^2-a^2}\left\{\frac{\partial}{\partial a^2}G(p)+
{\cal C_1}G(p)
\right\}+{\cal C}_2\frac{\partial}{\partial a^2}G(p)
\label{dGdaa}
\eeq
where the constants ${\cal C}_1$ and ${\cal C}_2$ are given by
\bea
{\cal C_1}&=&\frac{2}{3}\left\{
\frac{1}{4a^2}\rho_a\left(1-\rho_a\right)-\frac{1}{2}\frac{\partial
\rho_a}{\partial a^2}
-\frac{1}{2}\frac{1}{\lambda_a}(1-\rho_a)\frac{\partial \lambda_a}{
\partial a^2}\right\}
\label{c1} \\
{\cal C}_2 &=& \frac{1}{\lambda_a}\frac{\partial \lambda_a}{\partial a^2}.
\eea
We note that we always have a recursive relation like~\rf{dGdaa} relating the
$(k+1)^{th}$ derivative of $G(p)$ to the $k^{th}$ and the $(k-1)^{th}$.
This follows from the fact, already evoked several times, that any  solution of
the saddle point equation~\rf{saddle} can be parametrized in terms of any two
other independent solutions.
The nature of the parametrization follows from an analysis of the analyticity
structure and the asymptotic behaviour of the functions involved and a
recursive relation for the expansion coefficients can be found.
Since for the moment we will need only $\frac{\partial G(p)}{\partial a^2}$ and
$\frac{\partial^2 G(p)}{\partial (a^2)^2}$ we shall not enter into a detailed
discussion of this point, but we will make use of such considerations in
section \ref{higher} concerning the calculation of higher genera contributions.

Now, let us consider the action of the operator $\hat{K}$ on the functions
above.
One finds
\bea
\hat{K}\left(\frac{\partial G(p) }{\partial a^2}\right)&=&{\cal
M}_1\frac{1}{p^2-a^2},\\
\hat{K}\left(\frac{\partial^2 G(p)}{\partial (a^2)^2}\right)&=&
{\cal M}_2\frac{1}{p^2-a^2}+\frac{2}{3}{\cal M}_1\frac{1}{(p^2-a^2)^2}
\eea
where the moments ${\cal M}_1$ and ${\cal M}_2$ are defined by
\bea
{\cal M}_i&=&(4-n^2)\oint_{C_2}\frac{d\om}{2\pi i}\, \om
\left\{W_{r+}(\om)\left(\left(\frac{\partial}{\partial
a^2}\right)^iG_-(\omega)\right)+
W_{r-}(\om)\left(\left(\frac{\partial}{\partial
a^2}\right)^iG_+(\om)\right)\right\} \non \\
&=&2 \oint_{C_1}\frac{d\om}{2\pi i}\,\om\,
V'(\om)\left(\frac{\partial}{\partial a^2}
\right)^i G(\om), \hspace{0.7cm} i=1,2.
\eea
This means that we can choose our $\chi_a$-functions in the following way
\bea
\chi_a^{(1)}(p)&=&\frac{1}{\cal M_1}\frac{\partial}{\partial a^2} G(p)
\label{chi1a}  \\
\chi_a^{(2)}(p)&=&
\frac{1}{\cal M_1}\left\{\frac{2}{3}\left(\frac{\partial}{\partial
a^2}\right)^2
G(p)-\frac{\cal M_2}{\cal M_1}\frac{\partial}{\partial a^2} G(p)\right\}
\label{chi2a}
\eea
Needless to say that $\chi_b^{(m)}(p)$ appears from $\chi_a^{(m)}(p)$ by the
replacement $a\leftrightarrow b$.
Now combining the relations~\rf{A}, \rf{B} \rf{chi1a} and \rf{chi2a} one has an
explicit expression for the 1-loop correlator at genus one.

\subsection{The free energy at genus 1 \label{Fgenus1} }

To determine the free energy at genus one we use again the strategy of
reference~\cite{ACKM93}; namely we seek to express the basis vectors
$\chi_a^{(m)}(p)$ and $\chi_b^{(m)}(p)$ as total derivatives with respect to
the loop insertion operator.
The case $m=1$ is relatively simple.
Using the relation~\rf{dW-dV+} as well as~\rf{diffp}, \rf{diffa}, \rf{diffp2}
and~\rf{diffa2} one finds from the boundary equations~\rf{B1+} and~\rf{B2+}
after a lengthy but in principle straightforward calculation
\beq
\chi_a^{(1)}(p)=\frac{1}{4}\frac{d\log a^2}{dV(p)},\hspace{0.7cm}
\chi_b^{(1)}(p)=\frac{1}{4}\frac{d\log b^2}{dV(p)}.
\label{chi1ab}
\eeq
The case $m=2$ is less simple but due to the appearance of the factor ${\cal
M}_2/{\cal M}_1^2$ in the relation~\rf{chi2a} is is obvious that
$\chi_a^{(2)}(p)$ must be closely related to $d\log {\cal M}_1/dV(p)$.
By explicit computation one finds that this quantity can actually be expressed
entirely in terms of $\chi_a^2(p)$, $\chi_a^{(1)}(p)$, $\chi_b^{(1)}(p)$,
$a^2$, $b^2$ $e^2$, $\frac{\partial e^2}{\partial a^2}$ and $\frac{\partial
e^2}{\partial b^2}$ which is a non trivial result.
Let us briefly comment on the key relations which ensure this property.
(We will also need these relations for our discussions in
section~\ref{higher}.)
Acting with the loop insertion operator on ${\cal M}_1$ as usual implies
performing an explicit differentiation after the matrix model coupling
constants as well as an implicit differentiation after $a^2$ and $b^2$.
The explicit differentiation leads to the appearance of the quantity
$\frac{\partial}{\partial p}p \frac{\partial}{\partial a^2} G(p)$ which using
the relations~\rf{diffp},~\rf{diffa} and~\rf{diffa2} can be written as
\bea
\frac{\partial}{\partial p}\left(p\frac{\partial}{\partial a^2} G(p)\right)
&=&-2a^2\left(\frac{\partial}{\partial a^2}\right)^2G(p)
-b^2\left\{\frac{1}{b^2-a^2}+\frac{1}{e^2-b^2}\frac{\partial e^2}{\partial
b^2}\right\} \frac{\partial}{\partial b^2} G(p)\non \\
&& -\left\{\frac{2a^2-b^2}{a^2-b^2}+\frac{b^2}{e^2-a^2}\frac{\partial
e^2}{\partial b^2}\right\} \frac{\partial}{\partial a^2} G(p).
\label{explderiv}
\eea
The implicit differentiations lead to the appearance of mixed double
derivatives of $G(p)$ which with the use of~\rf{diffa} and~\rf{diffa2} can be
expressed in the following way
\bea
\frac{\partial}{\partial b^2}\frac{\partial}{\partial a^2} G(p) &=&
\frac{1}{2}\left\{\frac{1}{a^2-b^2}+\frac{1}{e^2-a^2}\frac{\partial e^2}
{\partial b^2}\right\}\frac{\partial}{\partial a^2}G(p) \non \\
&& +
\frac{1}{2}\left\{\frac{1}{b^2-a^2}+\frac{1}{e^2-b^2}\frac{\partial e^2}
{\partial a^2}\right\}\frac{\partial}{\partial b^2}G(p).
\label{mixderiv}
\eea
In total one ends up with the following expression for $\chi_a^{(2)}(p)$
\beq
3a^2 \chi_a^{(2)}(p)=-\frac{1}{2}\frac{d\log {\cal M}_1}{dV(p)}-
\frac{1}{4}\frac{d\log|a^2-b^2|}{dV(p)}-\frac{1}{2}\frac{d\log a^2}{dV(p)}
+\frac{1}{4} \frac{d\log|a^2-e^2|}{dV(p)}
\label{chi2dV}
\eeq
Evidently the relevant expression for $\chi_b^{(2)}(p)$ appears
from~\rf{chi2dV} by the interchangement $a^2\leftrightarrow b^2$.
Now inserting the here obtained expressions for the $\chi$-functions into the
expression~\rf{W1p} for $W^1(p)$ we can do the integration and obtain $F_1$.
The result reads
\bea
F_1&=&-\frac{1}{24}\log {\cal M}_1-\frac{1}{24}\log {\cal J}_1 -\frac{1}{6}\log
|a^2-b^2|
\non\\
&&+ \frac{1}{48}\log|a^2-e^2|+\frac{1}{48}\log a^2+\frac{1}{48}\log |b^2-e^2|+
\frac{1}{48}\log b^2 \non \\
&&+f_1\left(\frac{a}{b}\right)\label{F1}
\eea
where $f_1\left(\frac{a}{b}\right)$ obeys the following differential equation
\beq
xf_1'(x)=\frac{1}{4}\frac{e^2-\a^2}{a^2-b^2}
\eeq
and where ${\cal J}_1={\cal M}_1(a\leftrightarrow b)$.
We emphasize that this expression for $F_1$ holds for any potential $V(M)$ and
any $n\in ]-2,2[$.
The first three terms of~\rf{F1} have a structure similar to the terms which
appeared in the case of the 1-matrix model and one can easily verify that the
1-matrix model (n=0) result is correctly recovered.

\subsection{Higher genera and multi loops \label{higher} }

Having calculated $W^1(p)$ we are in a position to further iterate the genus
expanded version of the loop equation~\rf{loopgen}.
While the moments and basis vectors introduced in section~\ref{chidet}
certainly lead to simple expressions for the genus one quantities presented
there, they do not give the optimal parametrization of the model when it comes
to the representation of higher genera contributions.
Let us describe now what we consider as the optimal parametrization of the
model.
We will still work with a set of $\chi$-functions satisfying the
relations~\rf{eigenfunction} and~\rf{boundchi}.
However we will change the set of basis functions and moments.

As basis functions we shall use instead of the functions
$\left\{\left(\frac{{\partial}}{{\partial a^2}}\right)^kG(p),
\left(\frac{\partial}{\partial b^2}\right)^k G(p)\right\}$ a set of functions
$\{G_a^{(k)}(p),G_b^{(k)}(p)\}$ defined by
\label{areq}
\begin{enumerate}
\item $G_a^{(k)}(p)$ and $G_b^{(k)}(p)$ satisfy the homogeneous saddle point
equation~\rf{saddle}
$$ G^{(k)}(p+i0)+ G^{(k)}(p-i0) +n G^{(k)}(-p) =0,\hspace{0.7cm} p\in [a,b]. $$
\item $G_a^{(k)}(p)$ and $G_b^{(k)}(p)$ behave near the end points of the cut
$[a,b]$ as
$$G_{a}^{(k)}(p) \sim (p-a)^{-k-1/2}(p-b)^{-1/2}  \hs G_{b}^{(k)}(p) \sim
(p-b)^{-k-1/2}(p-a)^{-1/2}.  $$
\item $G_a^{(k)}(p)$ and $G_b^{(k)}(p)$ are analytical outside the cut
(especially near $-a$ and $-b$).
\item $G_a^{(k)}(p)$ and $G_b^{(k)}(p)$ have the following asymptotic behaviour
$$G_a^{(k)}(p),G_b^{(k)}(p) \sim \frac{1}{p^{k+1}}, \hspace{0.7cm} p\rightarrow
\infty .$$
\end{enumerate}
Here the conditions 1 and 3 ensure that the analyticity properties of
$G_a^{(k)}(p)$ and $G_b^{(k)}(p)$ are compatible with those of the one-loop
correlator and that $\hat{K}G_a^{(k)}(p)$ and $\hat{K}G_b^{(k)}(p)$ will be
even rational functions with poles at $p=\pm a$ and $p=\pm b$ respectively.
The purpose of condition 2 is simply to relate the degree of the poles to the
index $k$.
The conditions 1--3 are satisfied by many different families of functions, but
only one family of functions fulfills all four conditions.
We note that the set of functions $\left\{\left(\frac{\partial}{\partial
a^2}\right)^kG(p),\left(\frac{\partial}{\partial b^2}\right)^kG(p)\right\}$
introduced in section~\ref{1loop} fulfills the conditions 1--3 but not
condition 4.
Furthermore we note that for $n=0$ we reproduce exactly the basis functions
used in reference~\cite{ACKM93}.
Now for $n=2\cos(\nu \pi)$, let us denote by
$\left\{\tilde{G}^{(k)}_a(p),\tilde{G}^{(k)}_b(p)\right\}$ the basis functions
corresponding to $n=2\cos\left((1-\nu)\pi\right)$, i.e.
\beq
\left\{\tilde{G}^{(k)}_a(p),\tilde{G}_b^{(k)}(p)\right\}_{\nu}=
\left\{{G}^{(k)}_a(p),{G}_b^{(k)}(p)\right\}_{1-\nu}
\eeq
These are the functions that will appear in our definition of moments, namely
we define
\beq
M_k=2\oint_{C_1}\frac{d\om}{2\pi i}V'(\om)\tilde{G}^{(k)}_a(\om),
\hspace{1.0cm} J_k=M_k(a\leftrightarrow b).
\label{momentdef}
\eeq
In the case of the 1-matrix model (n=0) we have
$\left\{\tilde{G}^{(k)}_a(p),\tilde{G}_b^{(k)}(p)\right\}=
\left\{{G}^{(k)}_a(p),{G}_b^{(k)}(p)\right\}$ and hence we reproduce with our
definition exactly the moments (up to a factor 2) used for the 1-matrix
model~\cite{ACKM93}.
Furthermore as in the one-matrix model case, we can write one of the boundary
condition,~\rf{B1+}, as
\beq
M_0=0
\eeq
However, we stress that $n=0$ is a very special case.
In general we will have ${G}^{(k)}(p)\neq \tilde{G}^{(k)}(p)$.
We draw the attention of the reader to the importance of the condition number 4
in the definition of the basis functions.
With a boundary condition of this type we will have for a potential of degree
$d$ that $M_k=J_k=0$ for $k>d-1$.
This gives the parametrization of the model in terms of the smallest possible
number of moments\footnote{One may argue that we still have one parameter too
much since for a potential of degree $d$ described by $d$ coupling constants,
we have $d-1$ moment variables plus the two variables $a$ and $b$.
Indeed there is a constraint that would allow us to reduce the number of
parameters by one, namely the fact that the free energy has to be
dimensionless.
Hence, if we defined our moment variables to be dimensionless our results would
depend on $a$ and $b$ only via $a/b$.
However, we have not found that such a redefinition leads to any simplification
from a computational point of view.}.
The functions $\left\{\left(\frac{\partial}{\partial a^2}\right)^kG(p),
\left(\frac{\partial}{\partial b^2}\right)^k G(p)\right\}$ behave as ${\cal
O}(1/p^2)$ as $p\rightarrow \infty$.
Hence with these functions as basis functions we would have the unpleasant
situation that even with a potential of finite order we would have an infinite
number of moment variables.
We have kept these less pleasant moment variables in section~\ref{1loop} and
section~\ref{Fgenus1} since they render the formulation of the idea of our
iterative procedure more comprehensible and since they give a particularly
simple representation of the free energy for genus 1.
The nature of the prefactor of $1/p^{k+1}$ in the requirement on the asymptotic
behaviour of the basis functions is not important for the argument above.
However, it is convenient for the analysis of the critical behaviour of the
model that this prefactor is independent of $a$ and $b$.
We choose it equal to one for simplicity.
Let us mention that we can also write
\bea
M_k&=&(4-n^2)\oint_{C_2}\frac{d \om}{2\pi i} \left\{W_{r+}(\omega)
\tilde{G}_{a+}^{(k)}(\om)-W_{r-}(\om)\tilde{G}_{a-}^{(k)}(\om)\right\} \\
&=&(4-n^2)\oint_{C_2}\frac{d\om}{2\pi i}
\left\{W_{s+}(\om)\tilde{G}_{a+}^{(k)}(\omega)-W_{s-}(\om)\tilde{G}_{a-}^{(k)}
(\om)\right\}
\label{Mk+-}
\eea
The expression~\rf{Mk+-} is particularly appealing since the integrand does not
have any cut but only singularities in the form of poles at $\pm a$.
This follows from the fact that $W_{s\pm}(p)$ is a solution of the saddle point
equation~\rf{newsaddle} while $\tilde{G}_{a\pm}^{(k)}(p)$ is a solution of the
same equation with $\nu\rightarrow 1-\nu$ and that
$\tilde{G}_a^{(k)}(p)\sim(p-b)^{-1/2}$ for $p\sim b$ while $W_{s}(p)\sim
(p-b)^{1/2}$ for $p\sim b$.
Since in addition the integrand is odd, the contour $C_2$ can be deformed to a
small loop encircling the point $a$.
Similarly, in the case of the $J_k$ moments the contour $C_2$ can be deformed
into a small loop encircling the point $b$.
This observation will prove very useful for our considerations in
section~\ref{Oncrit} concerning the scaling limit of the model.

\subsubsection{Recursion relations for $G_a^{(k)}(p),G_b^{(k)}(p)$ }

{}From section~\ref{Gexpl} it follows that
\beq
G_a^{(0)}(p)=G_b^{(0)}(p)=\frac{G(p)}{2\cos(\nu\pi/2)}.
\eeq
Furthermore from the defining conditions 1--4 one can conclude
\beq
G_a^{(k+1)}(p)=\frac{1}{\lambda_a^{(k)}}\frac{\partial G_a^{(k)}(p)}{\partial
a^2}
\label{recursion1}
\eeq
since $\frac{\partial G_a^{(k)}(p)}{\partial a^2}$ fulfills the requirements
1--3 for $G_a^{(k+1)}(p)$ and the appropriate asymptotic behaviour can be
obtained by multiplication by a constant.
The expressions for $\lambda_a^{(0)}$ and $\lambda_a^{(1)}$ can be extracted
from the relations~\rf{diffa} and~\rf{c1} respectively.
One finds
\bea
\lambda_a^{(0)}&=&i\tan\left(\nu \pi/2\right)\lambda_a, \label{lambda0}\\
\lambda_a^{(0)} \lambda_a^{(1)}&=&-\frac{1}{2}\frac{\partial \rho_a}{\partial
a^2}
+\frac{1}{4a^2}\rho_a\left(1-\rho_a\right) \label{lambda1}
-\frac{1}{2}\frac{1}{\lambda_a}(1-\rho_a)\frac{\partial \lambda_a}{\partial
a^2}
\eea
We will now derive a set of recursion relations which allow us from the
knowledge of $G_a^{(0)}(p)$, $\lambda_a^{(0)}$ and $\lambda_a^{(1)}$ to
calculate any $G_a^{(k)}(p)$, $k>2$.
First we use the fact that any solution of the saddle point
equation~\rf{saddle} can be parametrized in terms of any two other independent
solutions to write
\beq
G_a^{(k+1)}(p)=\frac{1}{p^2-a^2}\left\{G_a^{(k-1)}(p)+
c_a^{(k)}G_a^{(k)}(p)\right\}.
\label{recursion2}
\eeq
Here the prefactor $1/(p^2-a^2)$ generates the correct leading singularity as
well as the correct asymptotic behaviour of $G_a^{(k+1)}(p)$.
The constant $c_a^{(k)}$ is determined by the requirement that $G_a^{(k+1)}(p)$
should not have a pole at $p=-a$, i.e.
\beq
c_a^{(k)}=-\frac{G_a^{(k-1)}(-a)}{G_a^{(k)}(-a)}.
\eeq
Next, by combining~\rf{recursion1} and~\rf{recursion2} we obtain the following
relations between the coefficients $c_a^{(k)}$ and $\lambda_a^{(k)}$
\beq
c_a^{(k)}\,\lambda_a^{(k)}=(k+\frac{1}{2}),\hspace{0.7cm}
\lambda_a^{(k+1)}-\lambda_a^{(k-1)}=\frac{\partial c_a^{(k)}}{\partial a^2}.
\label{clrecursion}
\eeq
{}From our knowledge of $G_a^{(0)}(p)$, $\lambda_a^{(0)}$ and $\lambda_a^{(1)}$
we can now by means of~\rf{recursion2} and~\rf{clrecursion} easily write down
an explicit expression for any $G_a^{(k)}(p)$ (and similarly for
$G_b^{(k)}(p)$).
Furthermore it is obvious that the $\tilde{G}$-functions appear from the
$G$-functions by the substitutions $\nu\rightarrow 1-\nu$ and we will use for
the relations involving $\tilde{G}$-functions the same notation as above just
with all quantities being equipes with a tilde.
We have in addition the following relation between the $G$ and $\tilde{G}$
functions
\beq
pG_a^{(k)}(p)=\tilde{G}_a^{(k-1)}(p)+s_a^{(k)}\tilde{G}_a^{(k)}(p)
\label{GGrel}
\eeq
The argument goes as above and the constant $s_a^{(k)}$ is given by
\beq
s_a^{(k)}=-\frac{\tilde{G}_a^{(k-1)}(0)}{\tilde{G}_a^{(k)}(0)}.
\label{gak}
\eeq
By inserting~\rf{GGrel} in~\rf{recursion1} and~\rf{recursion2} we find the
following expression for $s_a^{(k)}$
\beq
s_a^{(k+1)}=s_a^{(k)}\frac{\tilde{\lambda}_a^{(k)}}{\lambda_a^{(k)}}.
\label{srecursion}
\eeq
Hence it suffices to calculate $s_a^{(1)}$.
It reads
\beq
s_a^{(1)} =\frac{1}{2}\left(\frac{1-\rho_a}
{\lambda_a^{(0)}}\right).
\label{sa1}
\eeq
Now, if we define another set of moment variables by
\beq
\overline{M}_k=2\oint_{C_1}\frac{d\om}{2\pi i} \om V'(\om) G_a^{(k)}(\om),
\hspace{1.0cm} \overline{J}_k=\overline{M}_k(a\leftrightarrow b)
\label{Mkalt}
\eeq
we have
\beq
\overline{M}_k=M_{k-1}+s_a^{(k)}M_k
\label{MbarM}
\eeq
and inversely
\beq
M_k=\frac{1}{s_a^{(k)}}\overline{M}_k
-\frac{1}{s_a^{(k)}s_a^{(k-1)}}\overline{M}_{k-1}
+\ldots+(-1)^{k-1}\frac{1}{s_a^{(k)} \ldots s_a^{(1)}}\overline{M}_1.
\label{MMbar}
\eeq
To derive this equation one explicitly makes use of the fact that $M_0=0$.
These two relations allow us to move freely between the two sets of variables.
However, we stress that it is the $M$-moments which are the fundamental
quantities since these, as mentioned earlier, give the parametrization of the
model in terms of the smallest possible number of moments.
Working with the $\overline{M}$-moments would for a given potential of finite
degree (or for a given multi-critical point) lead to the appearance of one
additional parameter.
(For a potential of degree $d$ we will have $\overline{M}_k=0$ only for $k> d$
while $M_k$=0 for $k> d-1$.)

\subsubsection{Recursion relations for $\chi_a^{(k)}(p)$ and $\chi_b^{(k)}(p)$
}

We remind the reader that the aim of introducing the basis functions was to be
able to invert the operator $\hat{K}$.
Let us therefore examine the effect of acting with $\hat{K}$ on such a
function.
One finds
\beq
\hat{K}G_a^{(k)}(p)=\sum_{l=0}^{k-1}\frac{\mu_{k,l}}{(p^2-a^2)^{l+1}},
\hspace{0.7cm}
\hat{K}G_b^{(k)}(p)=\sum_{l=0}^{k-1}\frac{\tau_{k,l}}{(p^2-b^2)^{l+1}}
\label{KG}
\eeq
where $\mu_{k,l}$ and $\tau_{k,l}$ are defined by
\bea
\mu_{k,l} &=&(4-n^2)\oint_{C_2}\frac{d\om}{2\pi i} \om (\om^2-a^2)^l
\left\{W_{s+}(\om)G_{a-}^{(k)}(\om)+W_{s-}(\om)G_{a+}^{(k)}(\om)\right\}, \\
\tau_{k,l}&=&\mu_{k,l}(a\leftrightarrow b).
\eea
{}From~\rf{KG} we can write down a recursive relation for the $\chi$-functions,
namely
\beq
\chi_a^{(k)}(p)=\frac{1}{\mu_{k,k-1}}\left\{
G_a^{(k)}(p)-\sum_{i=1}^{k-1}\mu_{k,i-1}\chi_a^{(i)}(p)\right\}
\label{chia}
\eeq
and similarly for $\chi_b^{(k)}(p)$.
The $\mu$-coefficients can be expressed in  terms of the moment variables and
the $c_a^{(k)}$'s.
One has
\beq
\mu_{k,l}=0, \hspace{0.3cm} l\geq k,
\hspace{0.7cm} \mu_{k,0}=\overline{M}_k.
\eeq
and the remaining $\mu$-coefficients then follow from the recursion relation
\beq
\mu_{k+1,l}=\mu_{k-1,l-1}+c_a^{(k)}\mu_{k,l-1}
\eeq
which is a simple consequence of~\rf{recursion2}.
We note that in particular we have
\beq
\mu_{k,k-1}=\prod_{i=1}^{k-1}c_a^{(i)}\overline{M}_1
\eeq

\subsubsection{The one-loop correlator at genus $g$ }

In analogy with what was the case for the hermitian 1-matrix model we have the
following representation for the genus $g$ contribution to the 1-loop
correlator.
\beq
W^g(p)=\sum_{m=1}^{3g-1}\left\{A_g^{(m)}\chi_a^{(m)}(p)+B_g^{(m)}
\chi_b^{(m)}(p)\right\}
\label{Wg}
\eeq
where the $\chi$-functions are given by~\rf{chia} and where the coefficients
$A_g^{(m)}$ take the form
\beq
A_g^{(m)}=\sum f^{g,m}_{\beta_i,\gamma_j,\beta,\gamma}(a,b)\;
\frac{M_{\beta_1}\ldots M_{\beta_l}J_{\gamma_1}\ldots J_{\gamma_s}}{M_1^{\beta}
J_1^{\gamma}} \label{Agm}
\eeq
with the indices being restricted by the conditions
\bea
(l-\beta)+(s-\gamma)&=&2-2g \label{index1},\\
\sum_{i=1}^s(\beta_i-1)+\sum_{j=1}^l(\gamma_j-1)&\leq &3g-m-1. \label{index2}
\eea
That the equation~\rf{Wg} holds can be proven by induction using as the
starting point the expression obtained earlier for $W^1(p)$.
Obviously the proof consists in showing that with the representation~\rf{Wg}
valid for $g'=1,\ldots,g-1$ the right hand side of the loop
equation~\rf{loopgen} can be decomposed into fractions of the type
$(p^2-a^2)^{-m}$, $(p^2-b^2)^{-m}$, $m=1,\ldots,3g-1$ with appropriate
coefficients.
Let us just draw the attention of the reader to a few essential ingredients of
the proof.

As regards the first term on the right hand side of the loop equation the
existence of the above mentioned decomposition follows from the fact that the
basis functions fulfill the homogeneous saddle point equation and the
analyticity requirements 2 and 3 on page~\pageref{areq}.
This means that a function of the type $G_+^{(k)}(p)G_-^{(m)}(p)$ can not have
any cut but must be a rational fraction with poles at $p=\pm a$ and $p=\pm b$
of order less than or equal to $k+m$.

The important step in proving  that the second term on the right hand side
of~\rf{loopgen} indeed takes the desired form consists in showing that
$dM_k/dV(p)$ and $dJ_k/dV(p)$ can again be expressed in terms of basis
functions and $M$- and $J$-moments.
However, due to the relations~\rf{MbarM} and~\rf{MMbar} it is equivalent to
show that $d\overline{M}_k/dV(p)$ and $d\overline{J}_k/dV(p)$ can be expressed
in terms of basis functions and moments of the type $\overline{M_i}$,
$\overline{J_i}$.
For simplicity we shall here take the latter line of action.
{}From the definition~\rf{Mkalt} it follows that
\beq
\frac{d\overline{M}_k}{dV(p)}=2\frac{\partial}{\partial
p}\left(pG_a^{(k)}(p)\right)
+\frac{d a^2}{dV(p)}\lambda_a^{(k)}\overline{M}_{k+1}+
\frac{d b^2}{dV(p)}\,2\oint_{C_1}\frac{d\om}{2\pi i}\om V'(\om)\frac{\partial
G_a^{(k)}(\om)}{\partial b^2}
\label{dMdV1}
\eeq
where the first term comes from the explicit differentiation after coupling
constants and the two others from the implicit differentiation after $a^2$ and
$b^2$.
Exploiting the fact that $p^{k+1}G_a^{(k)}(p)$ depends on $p$ only via $p/a$
and $p/b$ we can rewrite the first term of~\rf{dMdV1} as
\beq
\frac{\partial}{\partial p}\left(pG_a^{(k)}(p)\right) =
-kG_a^{(k)}(p)-2a^2\frac{\partial G_a^{(k)}(p)}{\partial a^2}
-2b^2\frac{\partial G_a^{(k)}(p)}{\partial b^2}
\label{dpGdp}
\eeq
Furthermore the analyticity properties of $\frac{\partial
G_a^{(k)}(p)}{\partial b^2}$ allow us to conclude that we have a decomposition
of the following type
\beq
\frac{\partial G_a^{(k)}(p)}{\partial b^2}=v_{a,k}^{(0)}G_b^{(1)}(p)
+\sum_{i=1}^k v_{a,k}^{(i)}G_a^{(k)}(p)
\eeq
where the $v_{a,k}^{(i)}$'s are some constants.
{}From~\rf{mixderiv} it follows that for $k=1$ we have
\beq
v_{a,1}^{(0)}=-v_{a,1}^{(1)}=
\frac{1}{2}\frac{\lambda_b^{(0)}}{\lambda_a^{(0)}}\frac{\partial }{\partial
a^2}
\log \left(\frac{b^2-e^2}{b^2-a^2}\right)
\eeq
and the remaining $v$-coefficients can be found by repeatedly use of the $k=1$
relation and the relation~\rf{recursion1}.
In conclusion one can write $d\overline{M}_k/dV(p)$ as
\bea
\frac{d\overline{M}_k}{dV(p)}&=&
-2k G_a^{(k)}-4a^2
\lambda_a^{(k)}\left\{G_a^{(k+1)}(p)-\frac{\overline{M}_{k+1}}{\overline{M}_1}
G_a^{(1)}(p)\right\} \non \\
&&-4b^2\sum_{i=1}^k
v_{a,k}^{(i)}\left\{G_a^{(i)}(p)-\frac{\overline{M}_i}{\overline{J}_1}G_b^{(1)}(p)
\right\}
\label{dMdV}
\eea
where we note that the $v_{a,k}^{(0)}$ terms have cancelled.
Collecting the here given information it is straightforward to complete the
proof of the representation~\rf{Wg} for $W^g(p)$.
In case of the ordinary one-matrix model one has
$f_{\b_i,\gamma_j,\beta,\gamma}^{g,m}(a,b)=(a-b)^{-\delta}$ where
$\delta=4g-2-m-\sum_{i=1}^s(\beta_i-1)-\sum_{j=1}^l(\gamma_j-1)$.
In the general case this is no longer true.
However, we emphasize that we still have that all explicit dependence on the
matrix model coupling constants is hidden in the moment variables.
The function $f^{g,m}_{\beta_i,\gamma_j,\beta,\gamma}(a,b)$ is a function of
the endpoints of the cut only and expressed in terms of the variables $e$ and
$\alpha$ it takes the same form for all values of $n\in]-2,2[$.
Unfortunately we have not been able to write down the generic expression for
$f^{g,m}_{\beta_i,\gamma_i,\beta,\gamma}(e,\alpha)$.

\subsubsection{Multi-loop correlators}

 From $W_g(p)$ we can obtain $W_g(p_1,p_2,\ldots,p_s)$ for any $s$  by
repeatedly use of the loop insertion operator (cf.\ to
equation~\rf{multiloop}).
Analyzing the structure of the loop insertion operator, one can write down
formulas similar to~\rf{Wg} for the multi-loop correlators.
We will not pursue this aim, but let us mention that from the discussion in the
previous section it follow that the genus $g$ contribution to the $s$-loop
correlator as in the 1-matrix model case depends on at most $2\times (3g-2+s)$
moments for $g\geq 1$..
The same statement is true for $g=0$ provided $s\geq 3$.
This can be seen from the expression~\rf{W++pq} for the two-loop correlator at
genus zero.
We note that the expression~\rf{W++pq} could also have been obtained by
applying the loop insertion operator to the one-loop correlator at genus zero.
However, this method of calculation is more time consuming than the one
actually used.

\subsubsection{The free energy}

{}From $W^g(p)$ we can obtain $F_g$ by application of the inverse loop
insertion operator, the inversion being possible due to the relation~\rf{dMdV}.
One easily infers that as in the 1-matrix model case the genus $g$ contribution
to the free energy for $g\geq 1$ depends on at most $2\times(3g-2)$ moments and
that for $g\geq 2$, $F_g$ will be a sum of terms of the same type as those
entering the relation~\rf{Agm} where the indices fulfill~\rf{index1} as well as
a relation like~\rf{index2} where on the right hand side $3g-m-1$ is replaced
by $3g-3$.

\newsection{The cases $n=\pm 2$ \label{npm2} }

The cases $n=\pm 2$ pose no particular problems.
On the contrary they are in a certain sense easier to solve than the generic
cases, namely the saddle point equation as well as the loop equations can be
expressed in terms of functions of a definite parity and the generic solution
to the saddle point equation can be parametrized using only one singular
function.

\subsection{$n=-2$}

Let us start by noting that for $n=-2$, if we introduce $\La_i=\l_i^2$ in the
integral~\rf{eigenint} we find
\beq
Z \propto \int_0^{\infty} \prod_{i=1}^N d\La_i e^{-N\sum_i V(\sqrt{\La_i})}
\prod_{i<j}(\La_i-\La_j)^2
\eeq
Hence the partition function looks very similar to the one of the usual
hermitian 1-matrix model.
There are two important differences though.
Firstly the interval of integration is restricted to the positive real axis.
While this does not give rise to any complications concerning the solution
procedure it shows that the present model clearly contains other critical
points than the usual hermitian 1-matrix model;
namely points for which the eigenvalue distribution exactly touches the origin.
In this respect the model is very similar to the complex matrix model which is
given by an integral of the same type, the $\La_i$'s playing the role of the
positive eigenvalues of a matrix $\phi^{\dagger}\phi$~\cite{Morris}.
However, there is an important feature which differentiates the $O(-2)$ model
from both the complex and the hermitian one matrix model.
The potential $V(\sqrt{\La_i})$ might contain half integer powers of $\La_i$.
Likewise the correlation functions that one would be interested in calculating
will typical involve half integer powers of $\La_i$.
Let us proceed to discussing how the usual iterative procedure can be adjusted
to these circumstances.

\subsubsection{The one-loop correlator at genus zero \label{W0pn=-2} }

As in the previous sections we will assume that the 1-loop correlator $W(p)$
(defined by~\rf{multiloop}) is analytic in the complex plane and that it
behaves as $1/p$ as $p\rightarrow \infty$.
Let us decompose $W(p)$ as
\beq
W(p)=W_+(p)+pW_-(p) \label{oneloop}
\eeq
where the functions $W_+(p)$ and $W_-(p)$ are both even in $p$.
Now $W_+(p)$ and $W_-(p)$ have in addition to the cut $[a,b]$ a cut $[-b,-a]$
and the analyticity requirement on $W(p)$ implies
\beq
W_+(p+i0)-W_+(p-i0)=p\left(W_-(p+i0)-W_-(p-i0)\right),
\hspace{0.7cm}p\in [a,b]
\label{analyticity}
\eeq
In particular the eigenvalue density can be found from either one of the two
functions $W_+(p)$ and $W_-(p)$ (cf.\ to equation~\rf{density})
\bea
\rho(p)&=&\frac{p}{i\pi}\left(W_-(p+i0)-W_-(p-i0)\right),\hspace{0.6cm}
p\in[a,b] \non \\
&=&\frac{1}{i\pi}\left(W_+(p+i0)-W_+(p-i0)\right),\hspace{0.6cm}p\in[a,b]
\label{eigenvalue}
\eea
The saddle point equation~\rf{saddle} becomes an equation for $W_-(p)$ and
expressed in terms of the variable $p^2$ instead of $p$ it takes the same form
as the saddle point equation of the hermitian 1-matrix model.
Hence the solution of the present equation can be read off from the solution of
the latter.
One finds~\cite{KS92}
\beq
W_-(p)=\frac{1}{2}\oint_{C_1}\frac{d\om}{2\pi i} \frac{V'(\omega)}{p^2-\om^2}
\left\{\frac{(p^2-a^2)(p^2-b^2)}{(\om^2-a^2)(\om^2-b^2)}\right\}^{1/2}
\eeq
whew $a^2$ and $b^2$ are given by
\beq
\oint_{C_1}\frac{d\om}{2\pi i}
\frac{V'(\om)}{(\om^2-a^2)^{1/2}(\om^2-b^2)^{1/2}}=0,
\hspace{0.7cm}
\oint_{C_1}\frac{d\om}{2\pi i} \frac{V'(\om)\om^2
}{(\om^2-a^2)^{1/2}(\om^2-b^2)^{1/2}}=2
\label{b12}
\eeq
We note that from $W_-(p)$ we can find $W_+(p)$ by the following recipe
\beq
W_+(p)=2\oint_{C_1}\frac{d\om}{2\pi i} W_+(\om)\frac{\om}{p^2-\om^2}=
2\oint_{C_1}\frac{d\om}{2\pi i} W_-(\om)\frac{\om^2}{p^2-\om^2}
\eeq

\subsubsection{Higer genera and multi loops}

Let us introduce a decomposition of the loop insertion operator, namely
\beq
\frac{d}{dV(p)}=\frac{d}{dV_+(p)}+p\frac{d}{dV_-(p)}
\label{loopins}
\eeq
where the operators $d/dV_+(p)$ and $d/dV_-(p)$ contain only even powers of
$p$.
Then we can rewrite the loop equation~\rf{loopfinal} as
\beq
\frac{1}{p^2}\oint_{C_1}\frac{d\om}{2\pi i}\frac{V'(\om)\om^2}{p^2-\om^2}
W_-(\om)=\left(W_-(p)\right)^2 +\frac{1}{N^2}\frac{d}{dV_-(p)} W_-(p)
\label{loopn=-2}
\eeq
where we have explicitly made use of the relation~\rf{analyticity}.
Instead of searching a solution of the equation~\rf{loopn=-2} one can search a
solution of the following equation
\beq
\oint_{C_1}\frac{d\om}{2\pi i}\frac{V'(\om)}{p^2-\om^2}
W_-(\om)=\left(W_-(p)\right)^2 +\frac{1}{N^2}\frac{d}{dV_-(p)} W_-(p)
\label{loopn=-2A}
\eeq
since such a function will automatically fulfill
\beq
\oint_{C_1} d\omega V'(\om) W_-(\om)=0
\eeq
The genus $g$ contribution to the free energy of the $O(-2)$ model now takes
the same form as the genus $g$ contribution of the free energy of the hermitian
one-matrix model given in reference~\cite{ACKM93} provided the moments $M_k$
and $J_k$ are defined by
\bea
M_k&=&\oint_{C_1}\frac{d\om}{2\pi i}
\frac{V'(\omega)}{(\om^2-a^2)^{k+1/2}(\om^2-b^2)^{1/2}}\\
J_k&=& M_k(a^2\leftrightarrow b^2)
\eea
and the parameter $d$ is replaced by
\beq
d=b^2-a^2.
\eeq
This statement is easily proven.
First one rewrites the loop insertion operator $d/dV_-(p)$ in the moment
parametrization and realizes that it takes the same form as the loop insertion
operator of the hermitian one-matrix model (with the modifications given above)
except for $p$ being replaced by $p^2$.
This means that the analogy between the loop equations of the two models holds
to all orders in the genus expansion.
Secondly one notes that for the $O(-2)$ model the following obvious relation
holds
\beq
W^g_-(p)=\frac{d}{dV_-(p)}F_g
\eeq
and the correctness of the statement concerning the free energy becomes evident
after a few moments thoughts.

We emphasize that the contour $C_1$ above encircles only the cut $[a,b]$.
If the potential is even, however, one can immediately rewrite the integrals
above as integrals along the contour $C_2$.
Then performing the change of variable $\om^2 \rightarrow \omega$ one
reproduces exactly the expression for the free energy of the hermitian one
matrix model (of course with the assumption that the support of the eigenvalue
distribution lies on the positive real axis).

We will not pursue the explicit calculation of multi-loop correlators for the
$O(-2)$ model in the present publication but let us emphasize that such
calculations pose no particular difficulties.
One simply rewrites the loop insertion operator in the moment parametrization,
using the boundary equations \rf{b12} and applies it to the free energy.
As mentioned above $d/dV_-(p)$ has a structure similar to the loop insertion
operator of the hermitian  1-matrix model.
The even part $d/dV_+(p)$, however, is less simple and involves elliptic
integrals.

\subsection{$n=+2$}
\subsubsection{The one-loop correlator at genus zero \label{W0pn=2} }

Let us introduce again the decomposition  of the 1-loop correlator  given in
equation~\rf{oneloop}.
As before we then have the relation~\rf{analyticity} between $W_+(p)$ and
$W_-(p)$ and as before the eigenvalue density can be found from either of the
two as described in equation~\rf{eigenvalue}.
The saddle point equation turns into an equation for $W^0_+(p)$.
This equation when expressed in terms of $p^2$ takes the same form as the
saddle point equation for the hermitian 1-matrix model and the solution of the
present equation can be found by exploiting the analogy with the latter.
The result for $W_+^0(p)$ reads~\cite{KS92}
\beq
W_+^0(p)=\frac{1}{2}\oint_{C_1}\frac{d\om}{2\pi i}\frac{V'(\om)\om}{p^2-\om^2}
\left\{\frac{(p^2-a^2)(p^2-b^2)}{(\om^2-a^2)(\om^2-b^2)}\right\}^{1/2}
\label{solution}.
\eeq
Of the two boundary conditions which determine $a^2$ and $b^2$ one ensures the
correct asymptotic behaviour, $W_+(p)\sim O(1/p^2)$ as $p\rightarrow\infty$,
and can be written in the standard form
\beq
\oint_{C_1}\frac{d\om}{2\pi i}
\frac{V'(\om)\om}{(\om^2-a^2)^{1/2}(\om^2-b^2)^{1/2}}=0
\label{asymptotic}
\eeq
The other one expresses the fact that the eigenvalue distribution is normalized
to one and reads
\beq
\oint_{C_1}\frac{d\om}{2\pi i}\,W_+^0(\om)=\frac{1}{2} \label{B2n=2}
\eeq
or
\beq
\frac{1}{\pi}\int_a^b dp\, (p^2-a^2)^{1/2}(b^2-p^2)^{1/2}
\int_{C_1}\frac{d\om}{2\pi i}\frac{1}{p^2-\om^2}\frac{V'(\om)}
{(\om^2-a^2)^{1/2}(\om^2-b^2)^{1/2}}=2
\eeq
As opposed to what is normally the case this second condition can not be
written as a single contour integral.
This is due to the fact that $W_+(p)$ contains only the even powers of $p$,
i.e.\ the behaviour $W(p)\sim 1/p$ can not as usual be imposed by simply
referring to the contour integral~\rf{solution}.

Even though the complexity of the second boundary equation does render the
iterative calculation of the free energy and the multi-loop correlators more
involved than for $n=-2$, the moment technique is still applicable.
However, a detailed analysis of the structure of the free energy and the
multi-loop correlators at higher genera is rather work demanding and we shall
in the present publication restrict ourselves to exemplifying the applicability
of the moment description by calculating the free energy at genus 1.
Our line of action will follow closely the one taken for $n\in ]-2,2[$.

\subsubsection{The two-loop correlator at genus zero}

Introducing the decomposition~\rf{loopins} of the loop insertion operator we
can write the loop equation \rf{loopfinal} as
\beq
\oint_{C_1}\frac{d\omega}{2\pi i}\frac{\omega V'(\omega)}{p^2-\om^2}W_+(p)
=\left(W_+(p)\right)^2+\frac{1}{N^2}\frac{d}{dV_+(p)}W_+(p).
\label{loopn=2}
\eeq
which in its genus expanded version reads
\beq
\left\{\hat{K}-2W_+^0(p)\right\}W_+^g(p)=
\sum_{g'=1}^{g-1}W_+^{g'}(p)W_+^{g-g'}(p)+\frac{d}{dV_+(p)}W_+^{g-1}(p)
\label{loopgenn=2}
\eeq
where
\beq
\hat{K}f(p)=\oint_{C_1}\frac{d\om}{2\pi i}\frac{\om V'(\om)}{p^2-\om^2}f(\om)
\label{Khatn=2}
\eeq
To proceed with the solution we need to calculate the following two-loop
correlator
\beq
W^0_{++}(p,p)=\frac{d}{dV_+(p)}W^0_+(p)
\eeq
The simplest way to do this is to proceed as in section~\ref{twoloop}.
{}From~\rf{saddle} it follows that $W^0_{++}(p,q)$ must fulfill the following
saddle point equation
\beq
W_{++}^0(p+i0,q)+W_{++}^0(p-i0,q)=-\frac{1}{2}\frac{p^2+q^2}{(p^2-q^2)^2},
\hspace{0.7cm}p\in[a,b]
\eeq
The complete solution of this equation, consisting  of the sum of a particular
solution and the complete solution to the corresponding homogeneous equation,
is easily written down.
Using the fact that $W_{++}^0(p,q)$ must be symmetric in $p$ and $q$, finite
for $p=q$, have the asymptotic behaviour $W_{++}^0(p,q)\sim O(1/p^2)$ as
$p\rightarrow \infty$ and behave as $\left((p-a)(p-b)\right)^{1/2}$ in the
vicinity of $a$ and $b$.
one finds that it must necessarily take the form
\beq
W_{++}^0(p,q)=\frac{1}{8}\frac{1}{\overline{\sqrt{p}}}
\frac{1}{\overline{\sqrt{q}}}\left\{C+(p^2+q^2)\left(1-
\left(\frac{\overline{\sqrt{p}}-\overline{\sqrt{q}}}{p^2-q^2}\right)^2\right)
\right\} \label{W++n=2}
\eeq
where $C$ is some yet undetermined constant.
Now the boundary equation~\rf{B2n=2} implies that $W_{++}(p,q)$ must fulfill
the following equation
\beq
\oint_{C_1}dp\, W_{++}(p,q)=0,\hspace{0.7cm}\forall q
\eeq
{}From this equation one can extract the value of $C$.
This is most easily done evaluating the integral at $q=0$.
The result for $C$ reads
\beq
C=a^2+b^2-2b^2\frac{E(k_a)}{K(k_a)}=a^2+b^2-2a^2\frac{E(k_b)}{K(k_b)}
\eeq
where
\beq
k_a=\left(\frac{b^2-a^2}{a^2}\right)^{1/2},\hspace{0.7cm}
k_b=k_a(a^2\leftrightarrow b^2)
\eeq
and where $K(k_a)$ and $E(k_a)$ are the complete elliptic integrals of the
first and the second kind respectively.
To determine $W_{++}^0(p,p)$ which is the quantity which enters the loop
equation we must analyze carefully the limit $p\rightarrow q$ of the
expression~\rf{W++n=2}.
One finds
\beq
W_{++}^0(p,p)=\frac{1}{8}\frac{C}{(p^2-a^2)(p^2-b^2)}
+\frac{1}{16}\frac{p^2(a^2-b^2)^2}{(p^2-a^2)^2(p^2-b^2)^2}
\eeq
We note that the right hand side of the loop equation~\rf{loopgenn=2} for $g=1$
takes the same form as in the case $n\in]-2,2[\,\,$, the constant $C$ playing
the role of $e^2-\alpha^2$ (cf. to equation~\rf{W+-pp}).

\subsubsection{The one-loop correlator at genus one}

We shall try to express $W_+^1(p)$ as in equation~\rf{W1p} with the function
$\chi_a^{(i)}(p)$ and $\chi_b^{(i)}(p)$ obeying again the
relations~\rf{eigenfunction} and~\rf{boundary} with $\hat{K}$ given
by~\rf{Khatn=2}.
The corresponding $A$ and $B$ coefficients read
\bea
A_1^{(1)}=\frac{1}{8}\frac{1}{a^2-b^2}\left(C-\frac{1}{2}(a^2+b^2)\right),
&\hspace{0.6cm}& A_1^{(2)}=\frac{1}{16}a^2,
\label{An=2} \\
B_1^{(1)}=\frac{1}{8}\frac{1}{b^2-a^2}\left(C-\frac{1}{2}(a^2+b^2)\right),
&\hspace{0.6cm}& B_1^{(2)}=\frac{1}{16}b^2.
\label{Bn=2}
\eea
In analogy with the case  $n\in ]-2,2[$ we will express the $\chi$-functions in
terms of a set of basis functions $\{G_a^{(k)},G_b^{(k)}\}$.
To begin with let us introduce
\beq
\phi^{(o)}(p)=\frac{1}{(p^2-a^2)^{1/2}(p^2-b^2)^{1/2}}
\eeq
This function clearly fulfill the following identity
\beq
\hat{K}\phi^{(0)}(p)=0.
\eeq
 We now define $G_a^{(k)}(p)$ and $G_b^{(k)}(p)$ for $k\geq 1$ by the following
requirements
\begin{enumerate}
\item
$G_a^{(k)}(p)$ and $G_b^{(k)}(p)$ are even in $p$ and fulfill the homogeneous
saddle point equation
\beq
G(p+i0)+G(p-i0)=0,\hspace{0.7cm} p\in [a,b].
\nonumber
\eeq
\item
In the vicinity of the endpoints of the cuts the functions $G_a^{(k)}(p)$ and
$G_b^{(k)}(p)$ behave as
\beq
G_a^{(k)}(p)\sim(p^2-a^2)^{-k-1/2}(p^2-b^2)^{-1/2},
\hspace{0.7cm}
G_b^{(k)}(p)\sim(p^2-a^2)^{-1/2}(p^2-b^2)^{-k-1/2}.
\non
\eeq
\item
$G_a^{(k)}(p)$ and $G_b^{(k)}(p)$ are analytic everywhere else
\item
They fulfill the conditions
\beq
\oint_{C_1}dp\,G_a^{(k)}(p)=\oint_{C_1}dp\,G_b^{(k)}(p)=0. \non
\eeq
\item
They have the following asymptotic behaviour
\beq
G_a^{(k)}(p),G_b^{(k)}(p)=const\cdot \phi^{(0)}(p)+\frac{1}{p^{2k+2}}+
O\left(\frac{1}{p^{2k+4}}\right),\hspace{0.6cm} p\rightarrow \infty. \non
\eeq
\end{enumerate}
The role of the three first requirements is the same as in the generic case.
Condition number 4 ensures that the eigenvalue distribution stays normalized to
all orders in the genus expansion (cf. to equation~\rf{B2n=2}).
In the generic case this could simply be taken care of by demanding that
$W^g(p)\sim O(1/p^2)$, $p\rightarrow \infty$ for $g>1$, i.e. by excluding the
possibility of terms of order $1/p$ in the $G$-functions.
However, in the present case we are calculating only the even part of $W(p)$ so
the normalization condition must be imposed in a different way.
Condition 5 is chosen with the aim of rendering the moment variables as simple
as possible.
This should become clear shortly.
The conditions 1--5 determine the $G$-functions uniquely.
One has
\beq
G_a^{(k)}(p)=\phi_a^{(k)}(p)+S_a^{(k)} \phi^{(0)}(p)
\eeq
where
\beq
\phi_a^{(k)}(p)=\frac{1}{(p^2-a^2)^{k+1/2}(p^2-b^2)^{1/2}}
\eeq
and
\beq
S_a^{(k)}=-\frac{\oint_{C_1}d\om \,\phi_a^{(k)}(\om)}
{\oint_{C_1}d\om\,\phi^{(0)}(\om)}
\eeq
and similarly for $G_b^{(k)}(p)$. For our considerations in the following
section we will need the explicit expressions for $S_a^{(1)}$ and $S_a^{(2)}$.
They read
\bea
S_a^{(1)}&=&-\frac{1}{a^2-b^2}\left\{1-
\frac{E(k_b)}{K(k_b)}\right\}, \\
S_a^{(2)}&=&-\frac{1}{(a^2-b^2)^2a^2}\left\{a^2-\frac{1}{3}b^2+\frac{2}{3}
(b^2-2a^2)\frac{E(k_b)}{K(k_b)}\right\}.
\eea
{}From the basis functions it is straightforward to construct the
$\chi$-functions.
For that purpose let us consider the action of the operator $\hat{K}$ on the
$G$-functions.
One finds
\beq
\hat{K}G_a^{(k)}(p)=\sum_{l=1}^{k-1}M_l\frac{1}{(p^2-a^2)^l},
\hspace{1.0cm}
\hat{K}G_b^{(k)}(p)=\sum_{l=1}^{k-1}J_l\frac{1}{(p^2-b^2)^l}
\label{KGn=2}
\eeq
where the moments $M_l$ and $J_l$ are given by
\bea
M_k&=&\oint_{C_1}\frac{d\om}{2\pi i}\om V'(\om)G_a^{(k)}(\om)=
\oint_{C_1}\frac{d \om}{2\pi i}\frac{\om
V'(\om)}{(\om^2-a^2)^{k+1/2}(\om^2-b^2)^{1/2}} \\
J_k&=&M_k(a^2\leftrightarrow b^2)
\eea
The advantage of imposing the requirement 5 on the $G$-functions should be
clear by now.
One could have taken $G_a^{(k)}(p)$ as a linear combination of
$\phi^{(k)}_a(p)$ with any $\phi_a^{(l)}(p)$ with $l<k$.
However, due to the condition~\rf{asymptotic} we obtain a particularly simple
expression for the moments by choosing $l=0$. From~\rf{KGn=2} it follows that
the $\chi$-functions are given by
\beq
\chi_a^{(k)}(p)=\frac{1}{M_1}
\left\{G_a^{(k)}(p)-\sum_{l=1}^{k-1}M_{k-l+1}G_a^{(l)}(p)
\right\},\hspace{1.0cm} \chi_b^{(k)}(p)=\chi_a^{(k)}(p)\left(a\leftrightarrow
b\right)
\eeq
We note that for a potential of degree $p$ one has $M_q=J_q=0$ for $q>p$.
Now all the elements in the representation~\rf{W1p} of the 1-loop correlator at
genus one have been determined and it is easy, collecting the results of the
present section, to write down a completely explicit expression for $W_+^1(p)$.

\subsubsection{The free energy at genus one}

To determine the free energy at genus one we use the usual strategy of
expressing the $\chi$-functions as total derivatives with respect to the loop
insertion operator $d/dV_+(p)$.
The key point in this procedure consists in determining $da^2/dV_+(p)$ and
$db^2/dV_+(p)$.
These quantities can as usual be extracted from the boundary
conditions~\rf{asymptotic} and~\rf{normalization}.
The actual calculation is more involved than usual but after the use of various
relations between elliptic integrals one arrives at the following pleasant
result
\beq
\chi_a^{(1)}(p)=\frac{d\log a^2}{dV_+(p)},\hspace{1.0cm}
\chi_b^{(1)}(p)=\frac{d\log b^2}{dV_+(p)}
\eeq
Having obtained the expressions for $da^2/dV_+(p)$ and $db^2/dV_+(p)$ it is
relatively straightforward to show that
\beq
a^2\chi_a^{(2)}(p)=-\frac{2}{3}\frac{d\log M_1}{dV_+(p)}-\frac{1}{3}
\frac{d\log(a^2-b^2)}{dV_+(p)}-\frac{2}{3}\frac{d\log a^2}{dV_+(p)}
\eeq
and similarly for $\chi_b^{(2)}(p)$.
Now combining the $A$ and $B$ coefficients given in~\rf{An=2} and~\rf{Bn=2}
with the here obtained expressions for the $\chi$-functions one finds that
$W_+^1(p)$ indeed takes the form of a total derivative.
The free energy at genus one can hence be extracted and reads
\bea
F_1&=&-\frac{1}{24}\log M_1-\frac{1}{24}\log J_1-\frac{1}{6}\log(b^2-a^2)\non\\
&&+
\frac{1}{48}a^2+\frac{1}{48}\log b^2-\frac{1}{4}\log\left(K(k_a)\right)
-\frac{1}{4}\log\left(K(k_b)\right) \label{F1n=2}
\eea
It is interesting to note the similarity of~\rf{F1n=2} with the
expression~\rf{F1} obtained for $n\in]-2,2[$.

\newsection{The critical regime \label{Oncrit} }
\subsection{The critical points}

As mentioned earlier the matrix integral defining the $O(n)$ model ceases to
exist when the support of the eigenvalue distribution approaches zero, i.e.\
when $a\rightarrow 0$.
This gives rise to a new set of critical points for which no analogues exist
for the 1-matrix model~\cite{Kos89,KS92,EZ92}.
These are the critical points that we will consider in the following.
We will take $\nu$ to be in the interval $0<\nu<1$.
Then we allways have $a\ll |e| \ll b$ (since as we shall see very soon $e\sim
a^\nu$) which simplifies the analysis.
Although not more complicated the cases $\nu=0,1$  require special treatment.

At the singular points the eigenvalue distribution vanishes at one endpoint of
its support (here $a=0$) with a critical exponent, $\mu$, or equivalently
\beq
W_s(p)\sim p^{\mu}, \hspace{0.5cm} p\rightarrow 0.
\eeq
Let us recall the possible values of $\mu$ for the $O(n)$
model~\cite{KS92,EZ92}.
These can be read of from the expression~\rf{gensol} for $W_{s+}^{0}(p)$.
Obviously the possibility of new types of critical behaviour is due to the
presence of the function $G(p)$.
For $a=0$ (or equivalently $a\ll p \sim b$) the function $G(p)$ takes the form
\beq
G(p) \sim -\frac{2}{b}\cosh \psi \frac{\sinh(1-\nu)\psi}{\sinh
\psi},\hspace{0.7cm} \cosh \psi  =-\frac{b}{p}.
\label{G1(p)}
\eeq
This is most easily seen by verifying that the the function~\rf{G1(p)}
satisfies the criteria 1--3 on page~\pageref{Gcond}. Now letting $p\rightarrow
0$ we find that
\beq
G(p)\sim p^{\nu-1}+{\cal O}(p^{-\nu+1})\hspace{0.3cm}\mbox{and hence}
\hspace{0.3cm}\tilde{G}(p)\sim p^{-\nu}+{\cal O}(p^{\nu}).
\eeq
It then follows from~\rf{gensol}, \rf{Gtilde} and~\rf{invrot} that by fine
tuning the potential of our model (i.e.\ the polynomials ${\cal A}(p^2)$ and
${\cal B}(p^2)$) we can reach for a given value of $\nu$ (or $n$) the following
two series of critical points
\bea
\mu_{2m+1}=2m+1-\nu: &\hspace{0.7cm}& {\cal A}(p^2)\sim p^{2m},\hspace{0.4cm}
{\cal B}(p^2)\sim {\cal O}(p^{2m}) \\
\mu_{2m+2}=2m+1+\nu: &\hspace{0.7cm} &  {\cal B}(p^2)\sim p^{2m},
\hspace{0.4cm} {\cal A}(p^2)\sim {\cal O}(p^{2m+2})
\eea
The possible values of $\mu$ are exactly those for which $n=-2\cos(\mu\pi)$.
Furthermore it can be shown that
$\gamma_{str}=-2\nu/(\mu+1+\nu)$\cite{Kos89,KS92,EZ92}.
When $\nu=\frac{l}{q}$, with $0<l<q$ and $l,q\in Z$, the critical points being
characterized by the exponents $\mu=2m+1\pm \nu$ exhibit the scaling behaviour
characteristic of 2D gravity interacting with rational conformal matter fields
of the type $(q, (2m+1)q\pm l)$.
However, the continuum theories that one obtains from the $O(n)$ model do not
contain all the operators of the corresponding minimial
models~\cite{Kos89,KS92,EZ92}.

For later book-keeping purposes, let us arrange all critical points into one
series where the $M$'th multi-critical point is characterized by
\beq
\mu_M=M-\eta_{M+1},\hspace{1.2cm} \eta_{2k}=\nu,\hspace{0.5cm}
\eta_{2k+1}=1-\nu \label{muM}
\eeq
We note that this definition reproduces the usual notion of a $M$'th critical
point of the 1-matrix model ($\nu=\frac{1}{2}$) case.

\subsection{Scaling at a $M$'th critical point \label{scaling} }

In this section we will calculate the scaling behaviour of the basic elements
of our description, i.e.\ the functions $G_{a,b}^{(k)}(p)$ and the moments
$\{M_k,J_k\}$.
Knowing the scaling properties of these objects we can easily extract continuum
results from our exact results or develop a procedure for calculating directly
continuum quantities.

The most fundamental quantity of our description is the function $G(p)$.
{}From $G(p)$ all other quantities can be derived.
One can show that in the scaling region ($a\sim p\ll b$)
\beq
G(p)\sim -\frac{2ie}{ab} \frac{\sinh \nu \phi}{\sinh \phi},\hspace{0.7cm} \cosh
\phi =-\frac{p}{a}
\label{G2(p)}
\eeq
The prefactor comes from the relation~\rf{G+-}.
Now matching the  expressions~\rf{G1(p)} and~\rf{G2(p)} in the intermediate
region $a\ll p\ll b$ one can determine $e$ to leading order in $a$.
The result reads
\beq
e=2ib\left(\frac{a}{4b}\right)^{\nu},\hspace{0.4cm}\mbox{ i.e.} \hspace{1.0cm}
\rho_a=\nu
\eeq
which we note justifies our statements concerning $e$ made in
section~\ref{Gexpl}.
Hence for $p\sim a$ we have in accordance with the analysis of the previous
section
\beq
G(p)\sim a^{\nu-1},\hspace{1.0cm} \tilde{G}(p)\sim a^{-\nu}
\eeq
Knowing the scaling of $e$ we can furthermore determine the scaling of all
$G_a^{(k)}(p)$ and $G_b^{(k)}(p)$.
Namely, from the relations~\rf{lambda0} and~\rf{lambda1} we see that
$\lambda_a^{(0)}\sim a^{2\nu-2}$ and $\lambda_a^{(1)}\sim a^{-2\nu}$ and then
the recursion relations~\rf{clrecursion} tells us that
\beq
\lambda_a^{(k)}\sim \frac{1}{c_a^{(k)}}\sim a^{-2\eta_{k+1}}
\eeq
In particular $G_a^{(1)}(p)\sim a^{-1-\nu}$ and in general
\beq
G_a^{(k)}(p)\sim a^{-k-\eta_{k+1}},\hspace{0.7cm}
\tilde{G}_a^{(k)}\sim a^{-k-\eta_k},\hspace{1.0cm}p\sim a
\label{Gscal}
\eeq
while all the $G_b^{(k)}(p)$- and $\tilde{G}_b^{(k)}$-functions for $p\sim a$
become proportional to $G(p)$ and $\tilde{G}(p)$ respectively.

Let us now examine the scaling properties of the moment variables.
We remind the reader of the fact that the integrals defining $M_k$ and $J_k$
when written in the form~\rf{Mk+-} reduce to local integrations around $a$ and
$b$ respectively.
Since for $a\ll p\sim b$, $G^{(k)}(p)$ as well as $W_s(p)$ are independent of
$a$ we have
\beq
J_k\sim {\cal O}(a^0).
\eeq
The $M$-moments, on the contrary, have a non trivial scaling.
By definition of a $M$'th multi-critical point one has at such a point
\beq
W_s^{\mu_M}(p)\sim p^{M-\eta_{M+1}}\sim
a^{M-\eta_{M+1}}W_s^{\mu_M}\left(\frac{p}{a}\right).\hspace{0.7cm}
\eeq
It now follows that
\beq
M_k\sim a^{M-k+(\eta_{M}-\eta_k)}
\label{Mscal}
\eeq
We see that for $k<M$, $M_k$ scales with a positive power of $a$ and that
$M_M\sim a^0$.
The moments $M_k$ with $k>M$ are equal to zero.
This can be seen by deforming the contour to infinity.

Having determined the scaling properties of basis functions and moments it is
easy to pass to the continuum limit.
For instance to determine the genus one contribution to the 1-loop correlator
in the scaling limit it suffices to note that in this limit the right hand side
of the loop equations for $W^1(p)$ reduces to
\beq
\hat{K}W^1(p)=\frac{1}{4-n^2}\left\{\frac{a^2}{4}\frac{1}{(p^2-a^2)^2}-
\frac{2\nu^2 -4 \nu +1}{4} \frac{1}{(p^2-a^2)}\right\}.
\eeq
This immediately tells us that
\beq
W^1(p)=\frac{a^2}{4}\chi_a^{(2)}(p)-\frac{2\nu^2-4\nu+1}{4}\chi_a^{(1)}(p).
\label{W1cont}
\eeq
We note that there is no simplification of the $\chi$-functions in the scaling
limit.
All terms in the relation~\rf{chia} are of the same order in $a$.
Now using the relations~\rf{chi1a} and~\rf{chi2a} bearing in mind that ${\cal
M}_1=\lambda_a^{(0)}s_a^{(1)} M_1$ we find the following expression for $F_1$
in
the scaling limit.
\beq
F_1=-\frac{1}{24}\log\left(M_1 a^{-6\nu^2+13\nu-5}\right).
\label{F1cont}
\eeq
The exponent of $a$ vanishes if and only if $\nu=\frac{1}{2}$.
Hence we reproduce correctly the 1-matrix model result and we see once again
that this case is very particular.
{}From the expression~\rf{W1cont} of the scaling relevant part of $W^1(p)$ one
can pursue the iterative solution of the loop equation directly in the
continuum.
This only requires that one writes down a continuum  version of the loop
insertion operator.
Let us  mention a few properties of this operator.
First of all one finds that the loop insertion operator in the scaling limit
reduces to a differentiation after $a^2$ and the moments $M_k$.
This implies that, not surprisingly, no $J$-moments will appear in the scaling
limit.
Furthermore the dimension of the loop insertion operator can easily be
extracted.
It equals $a^{-\mu_M-2}$.
Let us stress that the expressions~\rf{W1cont} and~\rf{F1cont} as well as all
results that one would obtain by further iterations of the loop equations are
valid in the vicinity of any $M$'th critical point and independent of which
detailed prescription one might choose for approaching such a point.
However, whenever needed one can easily specialize to a given scaling
prescription.
In section~\ref{cosmological} we will show how one can calculate explicitly the
moments when one approaches the critical point by tuning an overall coupling
constant of the potential.

\subsection{The basis functions in the continuum \label{basis} }

As explained in the previous section the $G_b^{(k)}$-functions do not play any
role in the scaling limit.
Let us write the $G_a^{(k)}$-functions in this limit as
\beq
{G}_a^{(k)}(p)=\frac{1}{2\cos\left(\nu\pi/2\right) b^{k+1}}
\epsilon^{\nu_k}f_k(\phi)
\eeq
where we use again the parametrization $p=-a\cosh \phi$.
Now ${f}_k(\phi)$ is a dimensionless function and all dependence on the scaling
parameter $a$ is hidden in the prefactor $\epsilon^{{\nu}_k}$ where
\beq
\epsilon =\frac{a}{4b}.
\eeq
The value of the exponent ${\nu}_k$ follows from the relation~\rf{Gscal} and
reads
\beq
{\nu}_k=-k-\eta_{k+1}.
\eeq
The expression for the function ${f}_0(\phi)$ can be read off from~\rf{G2(p)}.
One has
\beq
{f}_0(\phi)=\frac{\sinh\nu\phi}{\sinh\phi}.
\eeq
The remaining ${f}_k$-functions can be found from the continuum versions of the
recursion relations~\rf{recursion1}, \rf{recursion2} and~\rf{clrecursion}.
They read
\bea
l_kf_{k+1}(\phi)&=&{\nu}_k{f}_k(\phi)-\frac{\cosh\phi}{\sinh\phi}  f_k'(\phi),
\label{ff'}\\
f_{k+1}(\phi)&=&\frac{1}{16\sinh^2\phi}\left(f_{k-1}(\phi)+\gamma_k f_k(\phi)
\right)
\eea
and
\beq
l_k\gamma_k=32(k+1/2),\hspace{1.0cm} l_{k+1}-l_{k-1}=2\gamma_k\eta_{k+1}.
\label{lgamma}
\eeq
The new dimensionless parameters $\gamma_k$ and $l_k$ are related to the
original ones $c_a^{(k)}$ and $\lambda_a^{(k)}$ by
\beq
\gamma_k=\lim_{a\rightarrow 0}\frac{c_a^{(k)}}{b}\epsilon^{-2\eta_{k+1}},
\hspace{1.0cm} l_k=\lim_{a\rightarrow 0}32b \lambda_a^{(k)}
\epsilon^{2\eta_{k+1}}.
\label{lgamma2}
\eeq
{}From the expressions~\rf{lambda0} and~\rf{lambda1} we can determine $l_0$ and
$l_1$ and this enables us to solve exactly the recursion relations~\rf{lgamma}.
We find
\beq
l_k=4\tan\left(\frac{\eta_k\pi}{2}\right)
\frac{\eta_{k}(1+\eta_{k})(2+\eta_{k})\ldots (k+\eta_k)}
{(1-\eta_k)(2-\eta_k)\ldots (k-\eta_k)}.
\label{lexplicit}
\eeq
One can derive additional interesting properties of the quantities appearing
above.
For instance one has
\beq
l_k l_{k+1}=16 \nu_k(\nu_k-1)
\eeq
and it appears that the $f$-functions satisfy the following differential
equation
\beq
f_k''(\phi)+2(k+1)\frac{\cosh \phi}{\sinh \phi} f_k'(\phi)+
\left((k+1)^2-\eta_{k}^2\right)f_k(\phi)=0.
\label{diffeq}
\eeq
As usual the relevant expressions for the $\tilde{G}$-functions appear from
those of the $G$-functions by the substitution $\nu\rightarrow 1-\nu$ and we
will use for the relations involving $\tilde{G}$ functions the same notation as
above just with all quantities being equiped with a tilde.
Equation~\rf{GGrel} relating $G$- and $\tilde{G}$-functions translates to the
scaling limit as
\beq
-4 \cosh\phi\, f_k(\phi) =\cot\left(\frac{\nu \pi}{2}\right)\left\{
\tilde{f}_{k-1}(\phi)+\sigma_k \tilde{f}_k(\phi)\right\}
\eeq
where the dimensionless parameter $\sigma_k$ is related to $s_a^{(k)}$ by
\beq
\sigma_k=\lim_{a\rightarrow 0}\epsilon^{-2\eta_k} \frac{1}{b} s_a^{(k)}.
\eeq
Using the relations~\rf{srecursion}, \rf{sa1}, \rf{lgamma2} and~\rf{lexplicit}
one can determine $\sigma_k$ explicitly.
It is given by
\beq
\sigma_k(k+\eta_k)=l_k.
\label{sigma}
\eeq

\subsection{Explicit calculations at a $M$'th multi critical point
\label{cosmological} }

In this section we specialize to a particular prescription for approaching a
$M$'th multi-critical point.
We replace the potential $V(p)$ of our model by $\frac{V_c(p)}{T}$ where
$V_c(p)$ is a critical potential corresponding to the critical point in
question and where $T$ plays the role of the cosmological constant (or the
temperature).
We now approach the critical point by letting $T\rightarrow T_c=1$ and define a
renormalized cosmological constant $\Lambda_R$ by~\cite{Kos89,KS92,EZ92}
\beq
T-T_c=a^{\mu_M+1-\nu} \Lambda_R.
\label{TTc}
\eeq
where $\mu_M=M-\eta_{M+1}$.
That the power of $a$ appearing above is indeed what is needed to make
$\Lambda_R$ dimensionless can be seen by expanding $W(p)$ around $W_c(p)$,
considering $p\sim a$ and using that
\beq
\frac{\partial\left(TW(p)\right)}{\partial T}=G_0(p).
\label{ddT}
\eeq
This relation follows from the fact that the expression on the left hand side
fulfills the conditions that determined uniquely the function $G_0(p)$.
Now, using the relation~\rf{TTc} it follows from~\rf{Mscal} and~\rf{F1cont}
that
\beq
F_1=-\frac{1}{24}\left\{1-\frac{6\nu^2-15\nu+7}{\mu_M+1-\nu}\right\}
\log\Lambda_R.
\eeq
This case is particularly simple.
Due to the logarithm we do not need to know the explicit expressions for the
moments in the scaling limit.
However, to determine the continuum version of any other quantity such
expressions are needed.
We shall now proceed to deriving these.
Our starting point will be the relation~\rf{Mk+-}.
We remind the reader that the contour integral appearing in this relation
reduces to a local integration around the point $a$.
Hence we only need to know the integrands in the scaling limit.
The relevant expressions for the $\tilde{G}$ functions appear from the previous
section.
However, we shall not make use of their explicit form. It suffices to know that
they fulfill the following relation with $x=p/a=-\cosh \phi$
\beq
\tilde{\nu}_k\tilde{f}_k(x)-x\tilde{f}_k'(x)=\tilde{l}_k\tilde{f}_{k+1}(x).
\label{ftilde}
\eeq
The scaling limit of the function $W_s(p)$ has been determined explicitly in
reference~\cite{EZ92}.
In the vicinity of a $M$'th multi-critical point one has
\beq
W_s(p)=\frac{1}{b}\,\epsilon^{M-\eta_{M+1}}F_M(x)
\eeq
where
\beq
F_M(x)=const\cdot\,\,\, x^{M-\eta_{M+1}}\left(
B+\int_{-\infty}^{\phi} d\alpha \frac{\sinh(\nu
\alpha)}{(\cosh\alpha)^{M-\eta_M}}\right)
\eeq
The constant $B$ can be completely explicited as a B Euler function but its
precise form will not be of importance for the following.
The prefactor is non universal and depends on the critical potential chosen.
{}From the explicit expression for $F_M(x)$ one easily verifies that the
following relation holds
\beq
(M-\eta_{M+1})F_M(x)-xF_M'(x)=const \cdot f_0(x)
\label{FM}
\eeq
Now our moments take the form
\beq
M_k=4\frac{\epsilon^{M+\eta_M+\tilde{\nu}_k}}{b^{k+1}} \hat{M}_k
\eeq
with
\beq
\hat{M}_k = (4-n^2)\oint \frac{dx}{2\pi i}
\left\{F_{M+}(x)\tilde{f}_{k+}(x)-F_{M-}(x)\tilde{f}_{k-}(x)\right\}
\equiv \langle F_M, \tilde{f}_k\rangle
\eeq
where the contour encircles the point $x=1$.
{}From the relations~\rf{FM} and~\rf{ftilde} it follows that
\beq
\tilde{l}_k\hat{M}_{k+1}=(M-k+\eta_M-\eta_k)\hat{M}_k- const\cdot
\langle f_0,\tilde{f}_k\rangle
\eeq
and since $\langle f_0,\tilde{f}_k\rangle\propto \delta_{k,0}$ we see that in
accordance with the analysis of section~\ref{scaling} the moments $M_k$ with
$k> M$ will vanish.
For $1\leq k\leq M$ we have
\beq
\frac{\hat{M}_{k+1}}{\hat{M}_k}=\frac{M-k+\eta_M-\eta_k}{\tilde{l}_k}
\label{MM}
\eeq
Hence we can express all our moments in terms of only one, say $M_1$.
This allows us to determine any continuum quantity up to a non-universal
constant.
For instance we find for $W_1(\phi)$ by means of~\rf{W1cont}, \rf{ff'},
\rf{sigma} and~\rf{MM}
\beq
W^1(\phi)=const\cdot \left\{l_1 f_2(\phi)-(\mu_M+1+6\nu^2-12\nu+3)f_1(\phi)
\right\}.
\eeq
Using the relations~\rf{ff'} and~\rf{diffeq} one can easily verify that this
results agrees with the one obtained in the unitary case within the framework
of strings with discrete target spaces by S.\ Higuchi and I.K.\
Kostov~\cite{HK95}.
\newsection{Conclusion and outlook}

One interesting conclusion which can be drawn from the obtained exact solution
of the $O(n)$ on a random lattice is that the model exhibits the same kind of
universality with respect to the potential as the hermitian 1-matrix model.
One must expect this kind of universality to occur also for two- and
multi-matrix models and the present work can be taken as an indicator of how
one could make use of this universality in the solution of these more
complicated models.

It is also interesting to note that our solution provides an exact solution of
the Ising model on a random lattice.
This gives the possibility of studying spin excitations of this model away from
criticality.
Unfortunately the representation of the Ising model on a random surface that
one obtains from the $O(n)$ model has vanishing magnetic field.
However, it is possible to include a magnetic field by adding a $1/M$ term to
the action appearing in equation~\rf{partition}.
In analogy with this one would expect that in general the addition of terms
with negative powers of $M$ would enlarge the operator content of the continuum
theories obtained from the model.
It would hence be interesting to generalize the moment technique to this
situation.

As mentioned in the introduction our d.s.l. relevant moment description of the
$O(n)$ model should allow us by comparison with the corresponding moment
description of the generalized Kontsevich models to determine which is the
precise relation between the continuum partition function of the  $O(n)$ model
for $n$ rational and the $\tau$-functions of the generalized kdV hierarchies.
We have not completed this analysis but let us mention a few observations.
First of all we see that for the $O(n)$ model on a random lattice we have in
the double scaling limit two series of moments with different scaling
properties.
In general for a $\tau$-function of the kdV$_p$ hierarchy describing the
interaction of 2D gravity with matter fields of the type
$(p,pm-1),\ldots,(p,pm-(p-1))$ there will appear $(p-1)$ series of moments with
different scaling properties~\cite{EYY94,KRI95}.
Hence the only models for which we could hope for an exact equivalence are the
models $(p,q)=(3,3m-1),$ $(3,3m-2)$.
However, as the example with the Ising model clearly shows, not even in this
case will the equivalence be exact.

Another interesting aspect concerning the double scaling limit is the
interpretation of the continuum theories corresponding to non-rational values
of $\nu$.
For instance, one might wonder what the topological interpretation of these
models is and if there exist integrable hierarchies describing them.

Finally one can remark that the results that we have obtained are actually
analytical in $\nu$.
This might open the possibility of attributing a meaning to the model for $n>2$
and maybe approaching the question of interaction of 2D gravity with matter
fields with $c>1$.\\

\vspace{12pt}
\noindent
{\bf Acknowledgements}\hspace{0.3cm}We thank I.\ Kostov for valuable
discussions and J.\ Zinn-Justin for providing us with the inspiration for the
solution of
the saddle point equation.

\end{document}